%% file: main.tex
\pdfoutput=1

\documentclass[12pt,a4paper]{article}

\usepackage{ifthen} 
\newboolean{pdflatex}
\setboolean{pdflatex}{true} 

\newboolean{articletitles}
\setboolean{articletitles}{true} 

\newboolean{uprightparticles}
\setboolean{uprightparticles}{false} 

\usepackage{graphicx}
\usepackage{pict2e}

\def\paperauthors{LHCb collaboration} 
\def\paperasciititle{Search for CP violation in Xic->pKpi decays using model-independent techniques} 
\def\papertitle{Search for \CP violation in \XicTopKpi decays using model-independent techniques} 
\def\paperkeywords{{High Energy Physics}, {LHCb}} 
\def\papercopyright{\the\year\ CERN for the benefit of the LHCb collaboration} 
\def\paperlicence{CC BY 4.0 licence}
\def\paperlicenceurl{https://creativecommons.org/licenses/by/4.0/}

\input{preamble}
\usepackage{longtable} 

\begin{document}

\renewcommand{\thefootnote}{\fnsymbol{footnote}}
\setcounter{footnote}{1}

\input{title-LHCb-PAPER}


\renewcommand{\thefootnote}{\arabic{footnote}}
\setcounter{footnote}{0}



\pagestyle{plain} 
\setcounter{page}{1}
\pagenumbering{arabic}


%

\input{introduction}

\input{selection}

\input{methods}

\input{control}

\input{results}

\input{summary}

\input{acknowledgements}




\clearpage

\addcontentsline{toc}{section}{References}
\bibliographystyle{LHCb}
\bibliography{main,standard,LHCb-PAPER,LHCb-CONF,LHCb-DP,LHCb-TDR}

\newpage
\input{LHCb_Authorship_25-Jun-2019.tex}

\end{document}

%% file: preamble.tex
\usepackage[top=1in, bottom=1.25in, left=1in, right=1in]{geometry}

\usepackage{microtype}
\usepackage{lineno}  
\usepackage{xspace} 
\usepackage{caption} 

\usepackage{graphicx}  
\usepackage{color}
\usepackage{colortbl}
\graphicspath{{./figs/}} 
\DeclareGraphicsExtensions{.pdf,.PDF,png,.PNG}

\usepackage{amsmath} 
\usepackage{amssymb}
\usepackage{amsfonts}
\usepackage{upgreek} 

\newcommand*\patchAmsMathEnvironmentForLineno[1]{%
\expandafter\let\csname old#1\expandafter\endcsname\csname #1\endcsname
\expandafter\let\csname oldend#1\expandafter\endcsname\csname
end#1\endcsname
 \renewenvironment{#1}%
   {\linenomath\csname old#1\endcsname}%
   {\csname oldend#1\endcsname\endlinenomath}%
}
\newcommand*\patchBothAmsMathEnvironmentsForLineno[1]{%
  \patchAmsMathEnvironmentForLineno{#1}%
  \patchAmsMathEnvironmentForLineno{#1*}%
}
\AtBeginDocument{%
\patchBothAmsMathEnvironmentsForLineno{equation}%
\patchBothAmsMathEnvironmentsForLineno{align}%
\patchBothAmsMathEnvironmentsForLineno{flalign}%
\patchBothAmsMathEnvironmentsForLineno{alignat}%
\patchBothAmsMathEnvironmentsForLineno{gather}%
\patchBothAmsMathEnvironmentsForLineno{multline}%
\patchBothAmsMathEnvironmentsForLineno{eqnarray}%
}

\usepackage{hyperxmp}

\usepackage[pdftex,
            pdfauthor={\paperauthors},
            pdftitle={\paperasciititle},
            pdfkeywords={\paperkeywords},
            pdfcopyright={Copyright (C) \papercopyright},
            pdflicenseurl={\paperlicenceurl}]{hyperref}

\usepackage[colorinlistoftodos,textsize=scriptsize]{todonotes}

\usepackage[all]{hypcap} 

\input{lhcb-symbols-def} 

\usepackage{cite} 
\usepackage{mciteplus}

%% file: lhcb-symbols-def.tex
\usepackage{xspace} 
\usepackage{upgreek}

\def\MagUp {\mbox{\em Mag\kern -0.05em Up}\xspace}

\ifthenelse{\boolean{uprightparticles}}%
{

 \def\Ppi         {\ensuremath{\uppi}\xspace}

 \def\PDelta      {\ensuremath{\Delta}\xspace}                 
 \def\PXi         {\ensuremath{\Xi}\xspace}                 
 \def\PLambda     {\ensuremath{\Lambda}\xspace}                 
 \def\PSigma      {\ensuremath{\Sigma}\xspace}                 
 \def\POmega      {\ensuremath{\Omega}\xspace}                 
 \def\PUpsilon    {\ensuremath{\Upsilon}\xspace}

 \def\PB      {\ensuremath{\mathrm{B}}\xspace}                 
                  
 \def\PD      {\ensuremath{\mathrm{D}}\xspace}

 \def\PK      {\ensuremath{\mathrm{K}}\xspace}

 \def\Pb      {\ensuremath{\mathrm{b}}\xspace}                 
 \def\Pc      {\ensuremath{\mathrm{c}}\xspace}

 \def\Pi      {\ensuremath{\mathrm{i}}\xspace}

 \def\Pp      {\ensuremath{\mathrm{p}}\xspace}

 \def\Ps      {\ensuremath{\mathrm{s}}\xspace}

 \def\thebaroffset{0.0em}
}
{

 \def\Ppi         {\ensuremath{\pi}\xspace}

 \mathchardef\PDelta="7101
 \mathchardef\PXi="7104
 \mathchardef\PLambda="7103
 \mathchardef\PSigma="7106
 \mathchardef\POmega="710A
 \mathchardef\PUpsilon="7107
                  
 \def\PB      {\ensuremath{B}\xspace}                 
                  
 \def\PD      {\ensuremath{D}\xspace}

 \def\PK      {\ensuremath{K}\xspace}

 \def\Pb      {\ensuremath{b}\xspace}                 
 \def\Pc      {\ensuremath{c}\xspace}

 \def\Pi      {\ensuremath{i}\xspace}

 \def\Pp      {\ensuremath{p}\xspace}

 \def\Ps      {\ensuremath{s}\xspace}

 \def\thebaroffset{0.18em}
}
\newcommand{\offsetoverline}[2][\thebaroffset]{\kern #1\overline{\kern -#1 #2}}%

\makeatletter
\ifcase \@ptsize \relax
  \newcommand{\miniscule}{\@setfontsize\miniscule{4}{5}}
\or
  \newcommand{\miniscule}{\@setfontsize\miniscule{5}{6}}
\or
  \newcommand{\miniscule}{\@setfontsize\miniscule{5}{6}}
\fi
\makeatother

\DeclareRobustCommand{\optbar}[1]{\shortstack{{\miniscule (\rule[.5ex]{1.25em}{.18mm})}
  \\ [-.7ex] $#1$}}

\def\squark    {{\ensuremath{\Ps}}\xspace}

\def\cquark    {{\ensuremath{\Pc}}\xspace}

\def\bquark    {{\ensuremath{\Pb}}\xspace}

\def\pion   {{\ensuremath{\Ppi}}\xspace}

\def\pip    {{\ensuremath{\pion^+}}\xspace}
\def\pim    {{\ensuremath{\pion^-}}\xspace}

\def\kaon    {{\ensuremath{\PK}}\xspace}

\def\KorKbar {\kern \thebaroffset\optbar{\kern -\thebaroffset \PK}{}\xspace}

\def\Kp      {{\ensuremath{\kaon^+}}\xspace}
\def\Km      {{\ensuremath{\kaon^-}}\xspace}

\def\DorDbar {\kern \thebaroffset\optbar{\kern -\thebaroffset \PD}\xspace}

\def\B       {{\ensuremath{\PB}}\xspace}

\def\BorBbar {\kern \thebaroffset\optbar{\kern -\thebaroffset \PB}\xspace}

\def\Bd      {{\ensuremath{\B^0}}\xspace}

\def\BdorBdbar {\kern \thebaroffset\optbar{\kern -\thebaroffset \Bd}\xspace}
\def\Bu      {{\ensuremath{\B^+}}\xspace}

\def\Bp      {{\ensuremath{\Bu}}\xspace}

\def\Bs      {{\ensuremath{\B^0_\squark}}\xspace}

\def\BsorBsbar {\kern \thebaroffset\optbar{\kern -\thebaroffset \Bs}\xspace}

\def\Y#1S{\ensuremath{\PUpsilon{(#1S)}}\xspace}

\def\proton      {{\ensuremath{\Pp}}\xspace}

\def\Deltares    {{\ensuremath{\PDelta}}\xspace}

\def\Lz          {{\ensuremath{\PLambda}}\xspace}

\def\LorLbar     {\kern \thebaroffset\optbar{\kern -\thebaroffset \PLambda}\xspace}

\def\Xires       {{\ensuremath{\PXi}}\xspace}

\def\Lc          {{\ensuremath{\Lz_\cquark}}\xspace}
\def\Lcp          {{\ensuremath{\Lz^+_\cquark}}\xspace}

\def\Xicp        {{\ensuremath{\Xires^+_\cquark}}\xspace}
\def\Xicm        {{\ensuremath{\Xires^-_\cquark}}\xspace}

\newcommand{\decay}[2]{\ensuremath{#1\!\to #2}\xspace}         

\def\to                 {\ensuremath{\rightarrow}\xspace}

\def\CP                {{\ensuremath{C\!P}}\xspace}
\def\CPV               {{\ensuremath{C\!PV}}\xspace}

\def\LcTopKpi     {\decay{\Lcp}{\proton\Km\pip}}
\def\XicTopKpi    {\decay{\Xicp}{\proton\Km\pip}}

\def\AT#1     {\ensuremath{A_{\mathrm{T}}^{#1}}\xspace}           

\def\C#1      {\ensuremath{\mathcal{C}_{#1}}\xspace}                       
\def\Cp#1     {\ensuremath{\mathcal{C}_{#1}^{'}}\xspace}                    
\def\Ceff#1   {\ensuremath{\mathcal{C}_{#1}^{\mathrm{(eff)}}}\xspace}        
\def\Cpeff#1  {\ensuremath{\mathcal{C}_{#1}^{'\mathrm{(eff)}}}\xspace}       
\def\Ope#1    {\ensuremath{\mathcal{O}_{#1}}\xspace}                       
\def\Opep#1   {\ensuremath{\mathcal{O}_{#1}^{'}}\xspace}                    

\newcommand{\aunit}[1]{\ensuremath{\text{\,#1}}}       

\newcommand{\tev}{\aunit{Te\kern -0.1em V}\xspace}
\newcommand{\gev}{\aunit{Ge\kern -0.1em V}\xspace}
\newcommand{\mev}{\aunit{Me\kern -0.1em V}\xspace}
\newcommand{\kev}{\aunit{ke\kern -0.1em V}\xspace}
\newcommand{\ev}{\aunit{e\kern -0.1em V}\xspace}
\newcommand{\mevc}{\ensuremath{\aunit{Me\kern -0.1em V\!/}c}\xspace}
\newcommand{\gevc}{\ensuremath{\aunit{Ge\kern -0.1em V\!/}c}\xspace}
\newcommand{\mevcc}{\ensuremath{\aunit{Me\kern -0.1em V\!/}c^2}\xspace}
\newcommand{\gevcc}{\ensuremath{\aunit{Ge\kern -0.1em V\!/}c^2}\xspace}

\def\fb   {\ensuremath{\aunit{fb}}\xspace}
\def\invfb   {\ensuremath{\fb^{-1}}\xspace}

\def\gsim{{~\raise.15em\hbox{$>$}\kern-.85em
          \lower.35em\hbox{$\sim$}~}\xspace}
\def\lsim{{~\raise.15em\hbox{$<$}\kern-.85em
          \lower.35em\hbox{$\sim$}~}\xspace}

\def\pt         {\ensuremath{p_{\mathrm{T}}}\xspace}

\xspace

\def\tell1  {TELL1\xspace}
\def\ukl1   {UKL1\xspace}



%% file: title-LHCb-PAPER.tex

\begin{titlepage}
\pagenumbering{roman}

\vspace*{-1.5cm}
\centerline{\large EUROPEAN ORGANIZATION FOR NUCLEAR RESEARCH (CERN)}
\vspace*{1.5cm}
\noindent
\begin{tabular*}{\linewidth}{lc@{\extracolsep{\fill}}r@{\extracolsep{0pt}}}
\ifthenelse{\boolean{pdflatex}}
{\vspace*{-1.5cm}\mbox{\!\!\!\includegraphics[width=.14\textwidth]{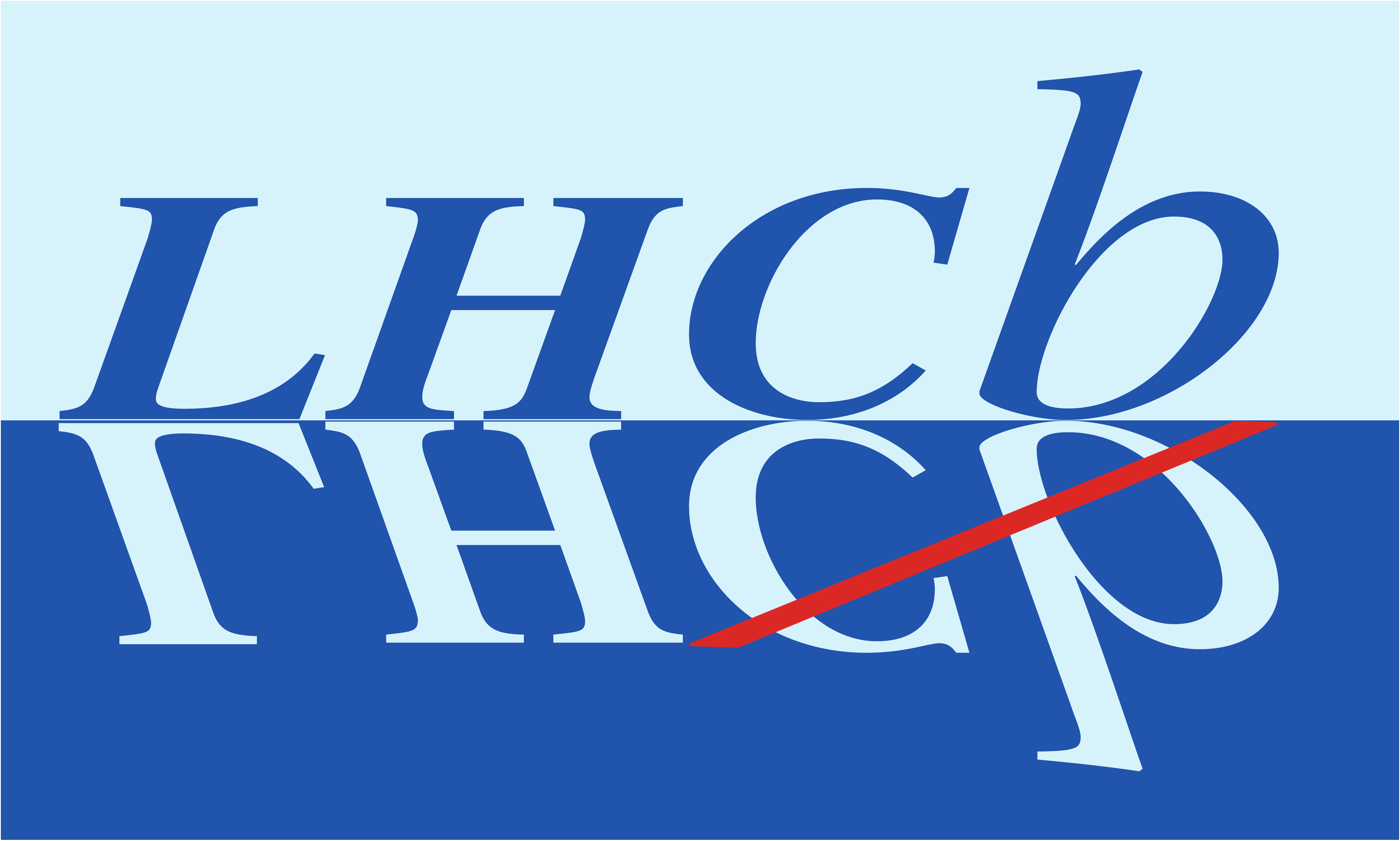}} & &}%
{\vspace*{-1.2cm}\mbox{\!\!\!\includegraphics[width=.12\textwidth]{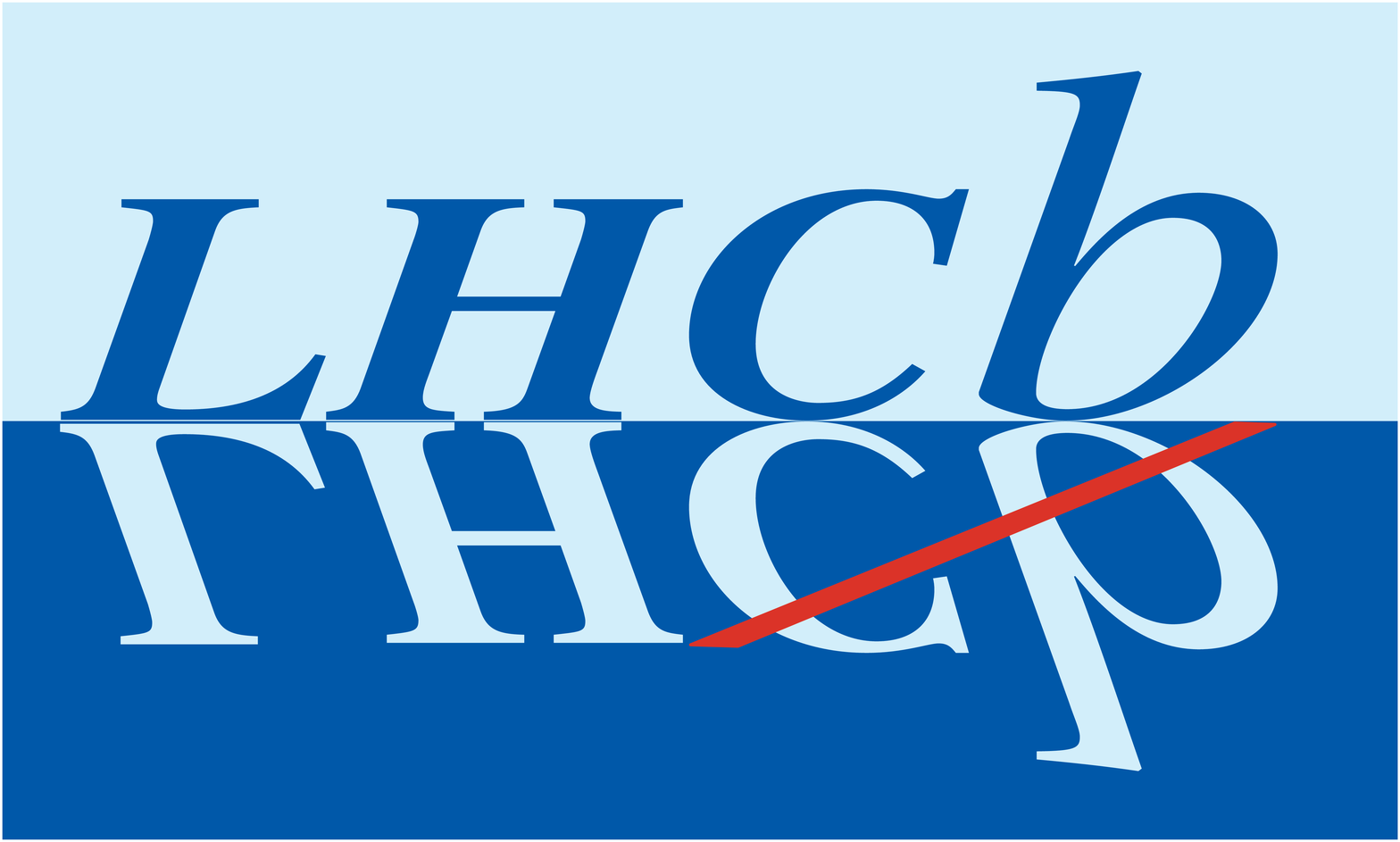}} & &}%
\\
 & & CERN-EP-2020-069 \\  
 & & LHCb-PAPER-2019-026 \\  
 & & 21 July 2020 \\
 & & \\
\end{tabular*}

\vspace*{4.0cm}

{\normalfont\bfseries\boldmath\huge
\begin{center}
  \papertitle 
\end{center}
}

\vspace*{2.0cm}

\begin{center}
\paperauthors\footnote{Authors are listed at the end of this paper.}
\end{center}

\vspace{\fill}

\begin{abstract}
  \noindent
 A first search for \CP violation in the Cabibbo-suppressed \mbox{\XicTopKpi} decay is performed using both a binned and an unbinned model-independent 
 technique in the Dalitz plot. The studies are based on a sample of proton-proton collision data, corresponding to an integrated luminosity of $3.0\invfb$, and collected by the LHCb experiment at centre-of-mass energies of $7$ and $8\tev$. The data are consistent with the hypothesis of no \CP violation.
  
\end{abstract}

\vspace*{2.0cm}

\begin{center}
  Submitted to
  Eur.~Phys.~J.~C 
\end{center}

\vspace{\fill}

{\footnotesize 
\centerline{\copyright~\papercopyright. \href{\paperlicenceurl}{\paperlicence}.}}
\vspace*{2mm}

\end{titlepage}


\newpage
\setcounter{page}{2}
\mbox{~}
%
%
%
%

\cleardoublepage

%% file: introduction.tex
\clearpage

\section{Introduction}

The non-invariance of fundamental interactions under the combination of charge conjugation and parity transformation, known as \CP violation (\CPV), is a key requirement for the generation of the baryon-antibaryon asymmetry 
in the early Universe~\cite{Sakharov:1967dj,Dine:2003ax}. 
In the Standard Model (SM) of particle physics, \CPV is included through the introduction of a single irreducible complex phase in the  Cabibbo--Kobayashi--Maskawa (CKM) quark-mixing 
matrix~\cite{Cabibbo:1963yz,Kobayashi:1973fv}. 
The amount of \CPV predicted by the CKM mechanism
is not sufficient to explain a matter-dominated
universe~\cite{Gavela:1993ts,Vieu:2018nfq}
and other sources of \CPV are required.
The realization of \CPV in nature has been well 
established in the $K$- and $B$-meson systems by several 
experiments~\cite{Christenson:1964fg,Aubert:2001nu,Abe:2001xe,Aubert:2004qm,Chao:2004mn,LHCb-PAPER-2013-018,LHCb-PAPER-2012-001}.
The LHCb experiment has observed for the first time \CPV
in the charm-meson sector as the difference 
of the \CP asymmetries between the two-body decays $D^0\rightarrow K^-K^+$ and 
$D^0\rightarrow \pi^-\pi^+$~\cite{LHCb-PAPER-2019-006}.
A similar study using \Lcp to \proton\Km\Kp
and \proton\pim\pip found no evidence for \CPV~\cite{LHCb-PAPER-2017-044}. Indeed, so far, \CPV has never been observed in any baryon system. Evidence for \CPV in the \bquark baryon sector reported by the LHCb collaboration in~\cite{LHCb-PAPER-2016-030}
has not been confirmed with more data~\cite{LHCb-PAPER-2019-028}.
Further measurements of processes involving the decay of charm hadrons can shed light on the origin and magnitude of \CPV mechanisms within the SM and beyond.

In two-body decays of charm hadrons, \CPV can manifest itself
as an asymmetry between partial decay rates.
Multi-body decays offer access to more observables 
that are
sensitive to \CP-violating effects. 
For a three-body baryon decay the kinematics can be characterised 
by three Euler angles and two squared invariant masses, which form 
a Dalitz plot~\cite{Zemach:1963bc}. 
The Euler angles are redundant if all initial spin states are integrated over. 
Interference effects
in the Dalitz plot probe \CP asymmetries in both 
the magnitudes and phases of amplitudes.
In three-body decays there can be
large local \CP asymmetries in
the Dalitz plot, even when no significant global
\CPV exists. A recent example has been measured in the
decay \decay{\Bp}{\pip\pim\pip}~\cite{LHCb-PAPER-2019-018}.

In the SM, \CPV asymmetries in the charm sector are expected at the order of $10^{-3}$ or less~\cite{BiancoBigi} for singly Cabibbo-suppressed (SCS) decays.
New physics (NP) contributions
can enhance \CP-violating effects
up to $10^{-2}$~\cite{Shipsey:2006zz,Artuso:2008CPV,Bianco:2020hzf,Grossman:2019xcj,Li:2019hho,Cheng:2019ggx,Calibbi:2019bay,Chala:2019fdb,Dery:2019ysp}.
Searches for \CPV in \Xicp baryon
decays\footnote{Unless stated explicitly, the inclusion of charge-conjugate states is implied throughout.}  
provide a test of the SM and place 
constraints on NP parameters~\cite{Grossman:2006jg,Bigi:2012ev,Grossman:2018ptn,Shi:2019vus,Wang:2019dls}.
 In contrast to SCS decays, in Cabibbo-favoured (CF) charm-quark transitions, such as \LcTopKpi decays, there is only one dominant amplitude in the SM, resulting in no \CP-violating effects. However this could change with NP, as argued above in the case of SCS decays.

This article describes searches for direct \CPV in 
the SCS decay \XicTopKpi, for $\Xicp$ baryons produced promptly
in \proton\proton collisions. 
The \LcTopKpi decay 
is used as a control mode
to study in data the level of experimental asymmetries that pollute the measurement.
In this paper, the symbol $H_{c}^+$ is used to refer to both \Xicp and \Lcp.
It is assumed that the polarisation of charm baryons 
produced in \proton\proton collisions is sufficiently 
small, as it is for \bquark-baryons~\cite{LHCb-PAPER-2020-005}, to justify the integration over the Euler angles.
This measurement uses \proton\proton collision data, corresponding 
to an integrated luminosity of $3\invfb$, recorded by the LHCb detector in 2011 ($1\invfb$) 
at a centre-of-mass 
energy of $7\tev$, and in 2012 ($2\invfb$)  at a
centre-of-mass energy of $8\tev$. 
The magnetic field polarity is reversed regularly during 
the data taking in order to minimise effects of charged particle 
and antiparticle detection asymmetries. 
Approximately half of the data are collected 
with each polarity.

There is presently no successful method for computing decay amplitudes in multi-body charm decays, which could provide reliable predictions on how the \CP asymmetries vary over the phase space of the decay. This situation favours a model-independent approach, which looks for differences between multivariate density distributions for baryons and antibaryons.
Therefore, in this article searches for \CPV are performed through a direct
comparison between the Dalitz plots of \Xicp and \Xicm decays using
a binned significance ($S_{\CP}$) method~\cite{miranda} and an unbinned 
k-nearest neighbour method (kNN)~\cite{Williams:2010vh,henze,schilling,LHCb-PAPER-2013-057},
both of which are model independent.


\section{Detector and simulation}

The LHCb detector~\cite{LHCb-DP-2008-001,LHCb-DP-2014-002} is a single-arm forward spectrometer 
covering the pseudorapidity range $2<\eta <5$. It is designed for the study
of particles containing $b$ and $c$ quarks. The detector includes
a high-precision tracking system consisting of a silicon-strip vertex
detector surrounding the $pp$ interaction region, a large-area 
silicon-strip detector located upstream of a dipole magnet
with a bending power of about $\rm 4~Tm$, and three stations
of silicon-strip detectors and straw drift tubes placed downstream of the magnet. 
The tracking system provides 
a measurement of the
momentum, $p$, of charged particles with a relative uncertainty that
varies from 0.5\% at low momentum to 1.0\% at $200~{\rm GeV}/c$.
The minimum distance of a track to a
primary vertex (PV),
the impact parameter (IP), is measured with a resolution of
$(15+ 29/\pt)~{\rm \mu m}$, where \pt is the component
of the momentum transverse to the beam, in ${\rm GeV}/c$.
Different types of charged hadrons are distinguished using
information from two ring-imaging Cherenkov detectors. Photons,
electrons and hadrons are identified by a calorimeter system
consisting of scintillating-pad and preshower detectors,
an electromagnetic and a hadron calorimeter. Muons are identified
by a system composed of alternating layers of iron and multiwire
proportional chambers. 

Samples of simulated events are used to optimise 
the signal selection, to derive the angular efficiency 
and to correct the decay-time efficiency.
In the simulation, \proton\proton 
collisions are generated using 
PYTHIA~\cite{Sjostrand:2007gs} with a specific LHCb
configuration~\cite{LHCb-PROC-2010-056}. Decays of hadronic
particles are described by EVTGEN~\cite{Lange:2001uf},
in which final-state radiation is generated using 
PHOTOS~\cite{Golonka:2005pn}. The interaction of the
generated particles with the detector, and its response, 
are implemented using the GEANT4 
toolkit~\cite{Allison:2006ve} as described in 
Ref.~\cite{LHCb-PROC-2011-006}.

%% file: selection.tex
\section{Selection of signal candidates}

The online event selection is performed by a trigger consisting
of a hardware stage, based on information from the calorimeter
and muon systems, followed by two software stages.
At the hardware trigger stage, 
events are required to have either 
muons with 
high \pt or hadrons, photons or electrons
with a high transverse-energy deposit in the
calorimeters. 
For hadrons, the transverse
energy threshold is approximately $3.5~{\rm GeV}/c^2$.
In the first software trigger stage at least one good-quality
track with $\pt>300~{\rm MeV}/c$ is required. 
In the second software trigger stage
an $H^+_c$ candidate is fully reconstructed 
 from three high-quality tracks not pointing to any PV. The three tracks should form a secondary vertex (SV) which must be well separated from any PV. 
 A momentum $p>3 \gevc$ for each track and the scalar sum of \pt for the three tracks $\pt>2 \gevc$ are required. The combined invariant mass of the three
 tracks is required to be in the range $2190 - 2570~\mevcc$. Requirements are also placed on the particle identification criteria of the tracks and on the angle between the vector from the associated PV to the SV and the $H_c^+$ momentum. The associated PV is
 the one with smallest difference in vertex fit $\chi^2$ when performed with and without the $H_c^+$ candidate.

In the offline analysis, 
tighter selection requirements are placed on 
the track-reconstruction quality, 
the $p$ and \pt of the final-state particles. For protons $10 <p < 100~\gevc$ is required, while kaons and pions momentum satisfies $3 <p< 150 \gevc$. Only $H_c^+$ candidates 
with \pt in the range $4 < \pt < 16~\gev$ are retained.
Additional requirements are also made
on the SV fit quality, and the minimum significance of the displacement from the SV to any PV in the event. This reduces the contribution
of charm baryons from \bquark-hadron decays to less than 5\%
of the prompt signal.
Reconstructed particles
are accepted if their momenta are within a region defined by
$|p_x|<0.2p_z$ and $|p_x|>0.01p_z$, where $p_x$ and
$p_z$ are the momentum components along the $x$ and $z$ axes\footnote{The LHCb coordinate system is right-handed, with the $z$ axis pointing along the
beam axis, $y$ the vertical direction, and $x$ the horizontal direction. The $(x, z)$ plane is the bending plane of the dipole magnet.}.
This requirement has a signal loss of 25\%, and is imposed to avoid large detection asymmetries that are present  in the excluded kinematic
regions. Differences between particles
and antiparticles in reconstruction efficiencies 
are also observed for $H^+_c$ candidates where
$p<20~{\rm GeV}/c$ for all charged 
tracks.
These differences do
not cancel by simply averaging the data acquired with opposite magnet polarities.
To minimise the reconstruction asymmetry, the momentum of all tracks is required to be greater than $20~{\rm GeV}/c$.
This requirement rejects about 20\% of the selected charm-baryon candidates.

\begin{figure}[!tb]
\begin{center}
\includegraphics[width=0.49\linewidth]{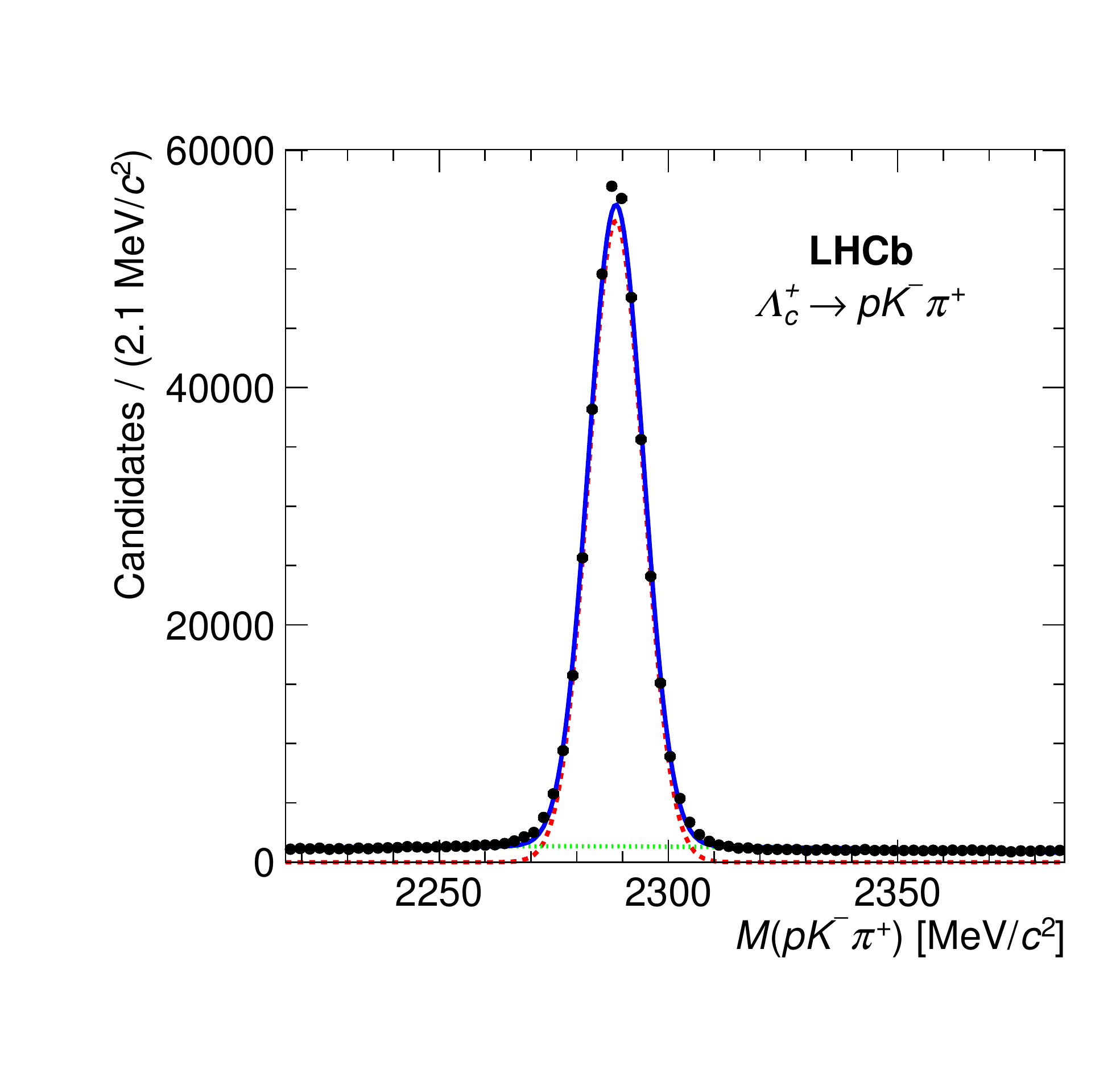}
\includegraphics[width=0.49\linewidth]{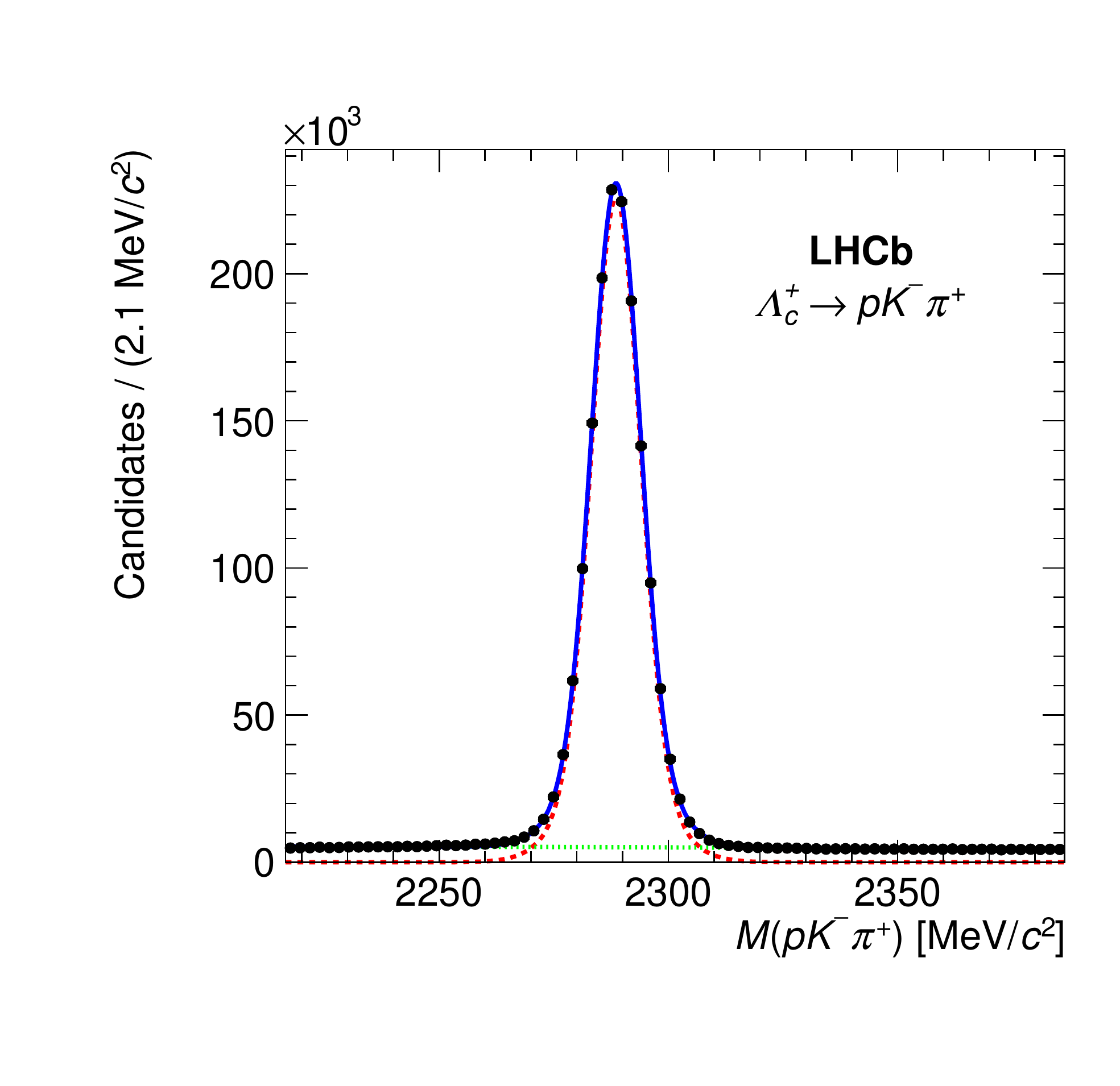}
\end{center}
\vspace*{-1.0cm}
\caption{Invariant-mass, $M(\proton\Km\pip)$,  distributions of selected
              \Lcp  candidates are shown in the (left) 2011 and (right) 2012 data samples.
              Data points are  in black. The overlaid fitted model (blue continuous line) is a sum 
              of two Gaussian functions with the same mean
              and different widths (red dashed line)
              and a second-order Chebyshev polynomial function
              (green dotted line) describing the signal and background components.}
\label{fig:massLambdafinal}
\end{figure}

\begin{figure}[!tb]
\begin{center}
\includegraphics[width=0.49\linewidth]{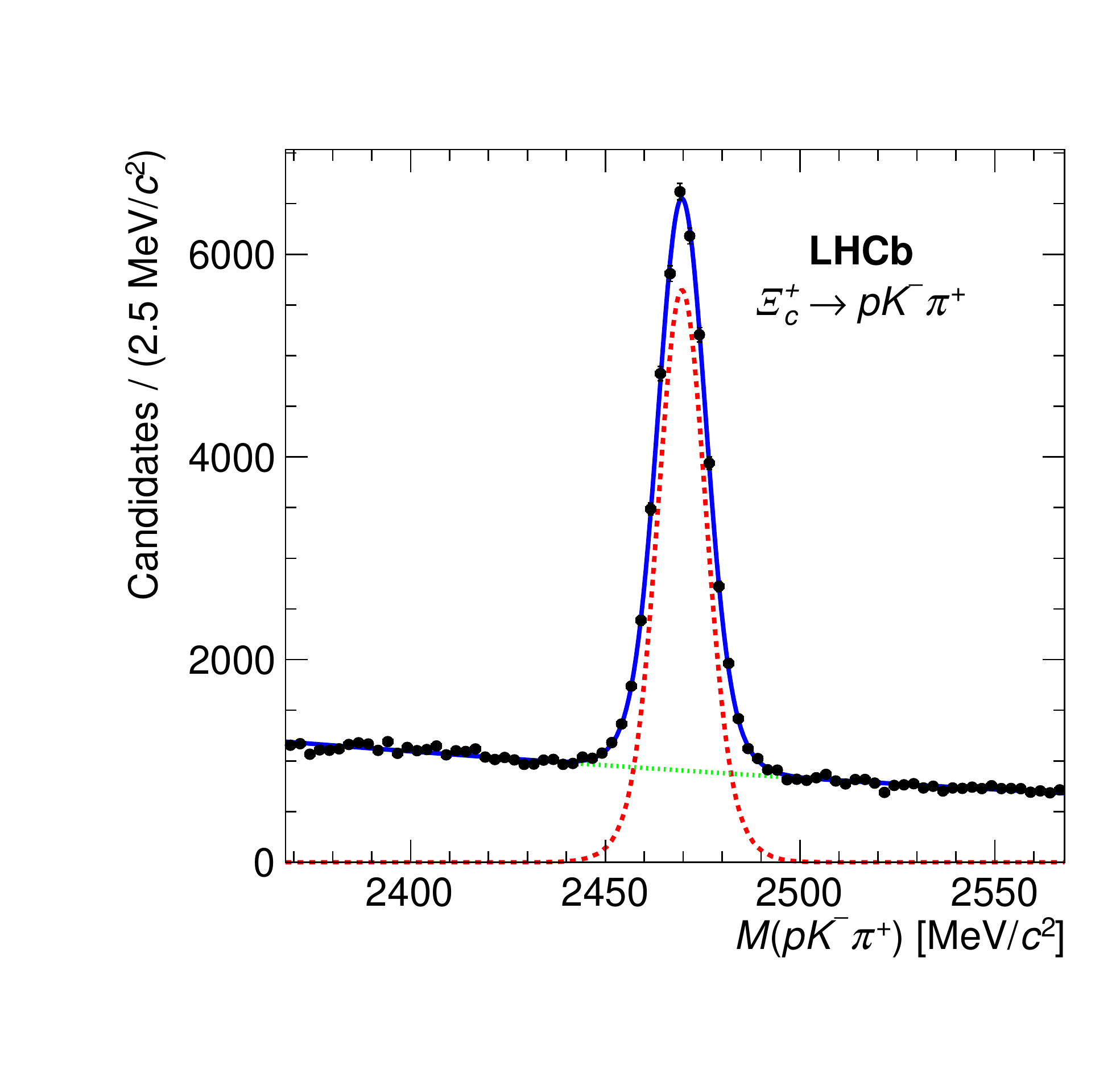}
\includegraphics[width=0.49\linewidth]{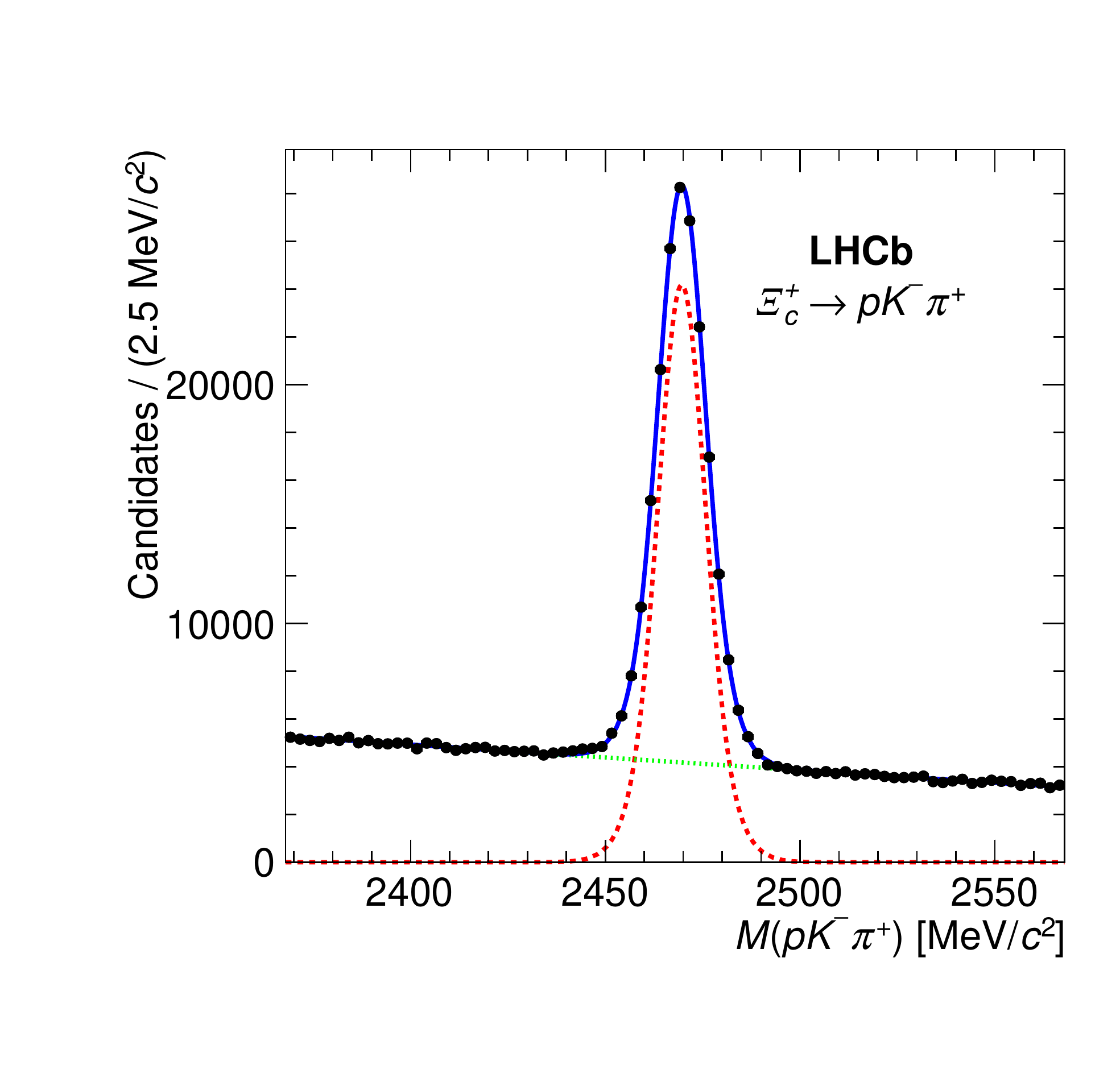}
\end{center}
\vspace*{-1.0cm}
\caption{Invariant-mass, $M(\proton\Km\pip)$, distributions of selected
              \Xicp  candidates are shown in the (left) 2011 and (right) 2012 data samples.
              Data points are  in black. The overlaid fitted model (blue continuous line) is a sum 
              of two Gaussian functions with the same mean
              and different widths (red dashed line)
              and a second-order Chebyshev polynomial function
              (green dotted line) describing the signal and background components.}
\label{fig:massXicfinal}
\end{figure}

The distributions of the invariant-mass,
$M(pK^-\pi^+)$, of selected \Lcp and \Xicp candidates are presented in Figs.~\ref{fig:massLambdafinal} and~\ref{fig:massXicfinal},
respectively, with fit curves overlaid. The fit model comprises
a sum of two Gaussian functions describing the signal and
a second-order Chebyshev polynomial function describing 
the combinatorial background. 
No additional source of
background is found to
contribute significantly, according to studies in data reconstructed with different mass hypotheses.

The final samples used for the \CPV search
comprise all candidates with
$M(pK^-\pi^+)$ within $\pm 3\sigma$
around $m(\Lcp)$ or $m(\Xicp)$, where
$\sigma$ is the weighted average of
the two fitted Gaussian widths
and $m(\Lcp)$ and $m(\Xicp)$
are the  masses of the \Lcp and \Xicp baryons~\cite{Tanabashi:2018oca}.
There are approximatly 2.0 million \Lcp candidates
(0.4 million  in the 2011 and  1.6 million  in the 2012 data sample) and 0.25 million \Xicp  candidates 
(0.05 million  in the 2011 and 0.2 million  in the 2012 data sample). 
The purity for \Lcp decays is 94\% for 2011 and 98\% for 2012
and that for \Xicp decays is 77\% for 2011 and 78\% for 2012, where purity is defined as the number of signal candidates  obtained from the fit to the invariant-mass distribution
divided by the total number of candidates.

%% file: methods.tex
\section{Methods}

\label{sec:methods}

The Dalitz plot for $H_{c}^+\to pK^-\pi^+$
is formed by the squares of the invariant masses of two pairs 
of the decay products:
$M^2(K^-\pi^+)$ and $M^2(pK^-)$.
Comparisons of the Dalitz plots of
$H_{c}^+$ and $H_{c}^-$ candidates are performed
using the binned $S_\CP$ and the unbinned kNN methods, 
described in the following.
For both the binned $S_{\CP}$ and unbinned kNN methods, 
a signal of \CPV is established if a {\it p}-value
lower then 
$3\times 10^{-7}$ is found, corresponding to an exclusion of \CP symmetry with a significance
of five standard deviations. 
However, in case that no \CPV is found, 
there is no model-independent mechanism for setting 
an upper limit on the amount of \CPV in the Dalitz plot.

\subsection{Binned \boldmath $S_\CP$ method}

\label{sec:methodsscp}

The $S_\CP$ method~\cite{miranda} has been
used before for searches of \CPV  testing in charm and beauty
decays~\cite{Aubert:2005gj,Aubert:2008yd,LHCb-PAPER-2011-017,LHCb-PAPER-2013-057,LHCb-PAPER-2014-020}.
This method is used to search
for localised asymmetries in the phase space of the 
decay 
$H_{c}^+\to pK^-\pi^+$  and
is based on a bin-by-bin comparison
between the Dalitz plots of baryons, $H_{c}^+$, and
antibaryons, $H_{c}^-$. 
The Dalitz plots 
of $H_{c}^+$ and $H_{c}^-$ are divided
using an identical binning. 
For each bin $i$ of the Dalitz plot,
the significance of the difference
between the number of $H_{c}^+$ ($n^i_+$) and 
$H_{c}^-$ ($n^i_-$) candidates,
is computed as

\begin{equation}
S^i_\CP = \frac{n^i_+ - \alpha n^i_-}{\sqrt{\alpha(n^i_+ + n^i_-)}},
\label{eq:scp}
\end{equation}
where the factor $\alpha$ is defined as 
$\alpha=\frac{n_+}{n_-}$ and $n_+$, $n_-$ are the total number of $H_{c}^+$, $H_{c}^-$ candidates.
This factor accounts for asymmetries arising in the  production of $H_{c}^+$
baryons, as well as in the detection of the final-state particles. The production 
and global detection asymmetries do not to depend on the Dalitz plot position.

A numerical comparison between the
Dalitz plots of the $H_c^+$ and $H_c^-$
candidates is made using a $\chi^2$ test defined as

\begin{equation}
\chi^2\equiv \Sigma (S^i_\CP)^2.
\end{equation}

A {\it p}-value for the hypothesis
of no \CPV is obtained from the $\chi^2$ distribution considering
that the number of degrees of freedom
is equal to the total number of bins minus
one, due to the constraint on the factor $\alpha$ of the overall $H_c^+$ and $H_c^-$ normalisation.

In the hypothesis of no \CPV, the $S_\CP$ values 
are expected to be distributed according to the normal distribution with a mean of zero and a standard deviation of unity. The test is performed using only bins with a minimum
of 10  $H_c^+$ and  10 $H_c^-$ candidates.
In case of \CPV, a deviation from the normal distribution is expected, generating a $p$-value close to zero.


\subsection{Unbinned kNN method}

\label{sec:methodsknn}

The kNN method is  based on the concept of a set of 
nearest neighbour candidates ($n_k$) 
in a combined sample of two data sets: baryons and
antibaryons. As an unbinned method, the kNN 
approach is more sensitive to a \CPV search
in a sample with limited data, 
compared to that of the binned $S_{\CP}$ method.
The kNN method is used here to test 
whether baryons and antibaryons share
the same parent distribution function~\cite{Williams:2010vh,henze,schilling}.
To find the $n_k$ nearest neighbour 
events 
of each $H_c^+$ or $H_c^-$ candidate,
an Euclidean distance between closest points in the Dalitz plot is used.
A test statistic $T$ for the null hypothesis
is defined as
\begin{equation}
T = \frac{1}{n_k(n_++n_-)}\sum\limits_{i=1}^{n_++n_-}\sum\limits_{k=1}^{n_k}I(i,k),
\label{eq:testT}
\end{equation}
where  $I(i,k)=1$ if the $i^{th}$ candidate and its $k^{th}$ nearest neighbour have the same charge
and $I(i,k)=0$ otherwise.

The test statistic $T$ is the mean fraction of like-charged neighbour pairs
in the sample of $H_c^+$ and $H_c^-$ decays. 
The advantage of the kNN method, 
in comparison with other proposed methods 
for unbinned analyses~\cite{Williams:2010vh}, 
is that the calculation of $T$ is simple 
and fast and the expected distribution 
of $T$ is well known.
Under the hypothesis of no \CPV, $T$ follows a normal distribution with a mean, $\mu_T$, and 
a variance, $\sigma_T$, where 
\begin{equation}
\mu_T = \frac{n_+(n_+-1)+n_-(n_--1)}{n(n-1)},
\label{eq:mut}
\end{equation}
\begin{equation}
\lim_{n,n_k,D \to \infty} \sigma_T^2 = \frac{1}{nn_k}\left(\frac{n_+n_-}{n^2}+4\frac{n_+^2n_-^2}{n^4}\right),
\label{eq:sigma}
\end{equation}
with $n=n_+ + n_-$ and $D=2$ is the dimensionality of the tested distribution. A good approximation of $\sigma_T$ is obtained even 
for $D = 2$ for the current values of $n_+$, $n_-$ and 
$n_k$~\cite{Williams:2010vh}.

For $n_+ = n_-$ the mean $\mu_T$ can be expressed as
\begin{equation}
\mu_{TR} = \frac{1}{2} \left( \frac{n-2}{n-1} \right)
\label{eq:muthalf}
\end{equation}
and is called the reference value, $\mu_{TR}$.
For large $n$, $\mu_{TR}$
asymptotically tends to $0.5$.

\begin{figure}[!htb]
\begin{center}
\includegraphics[width=0.49\linewidth]{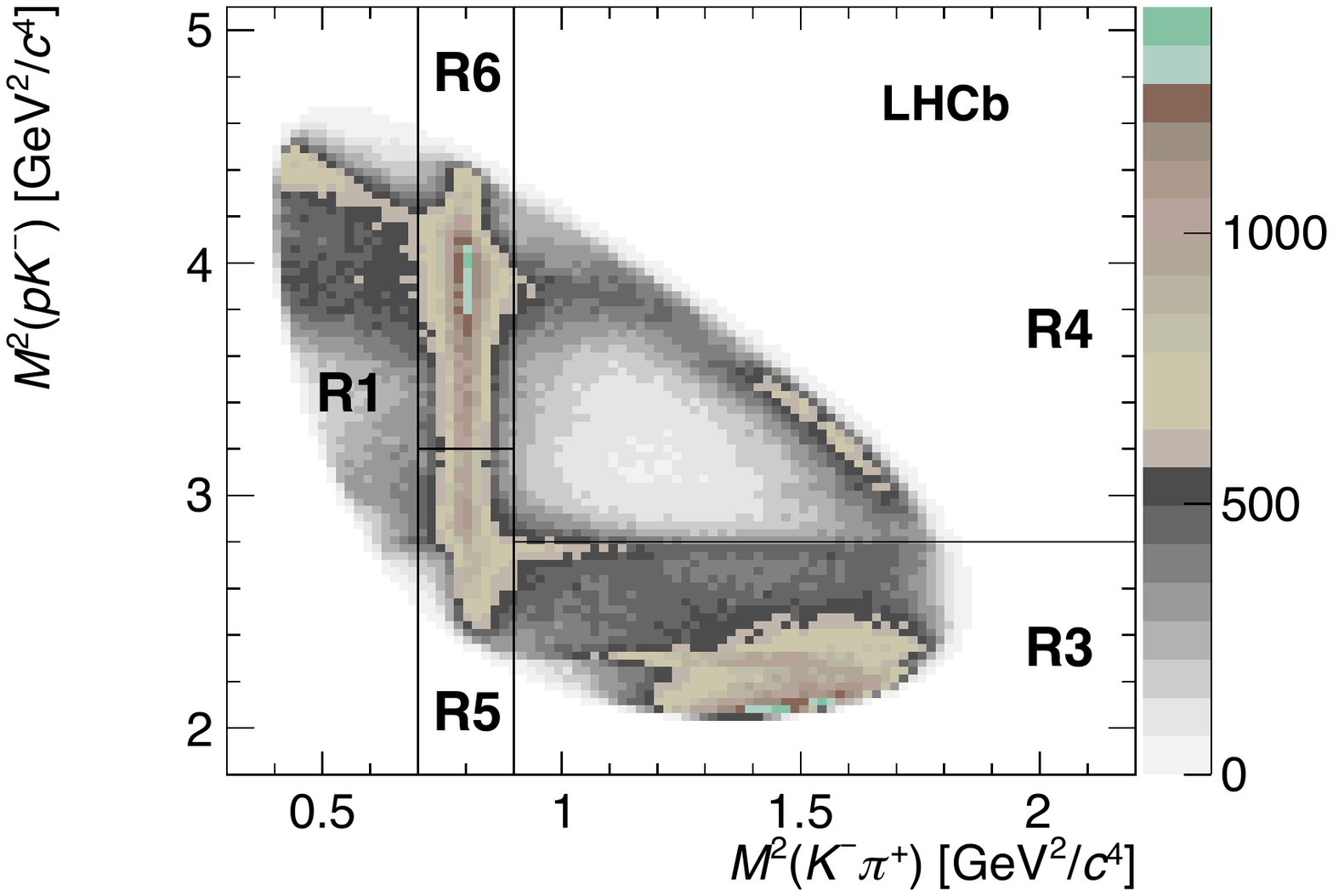}
\includegraphics[width=0.49\linewidth]{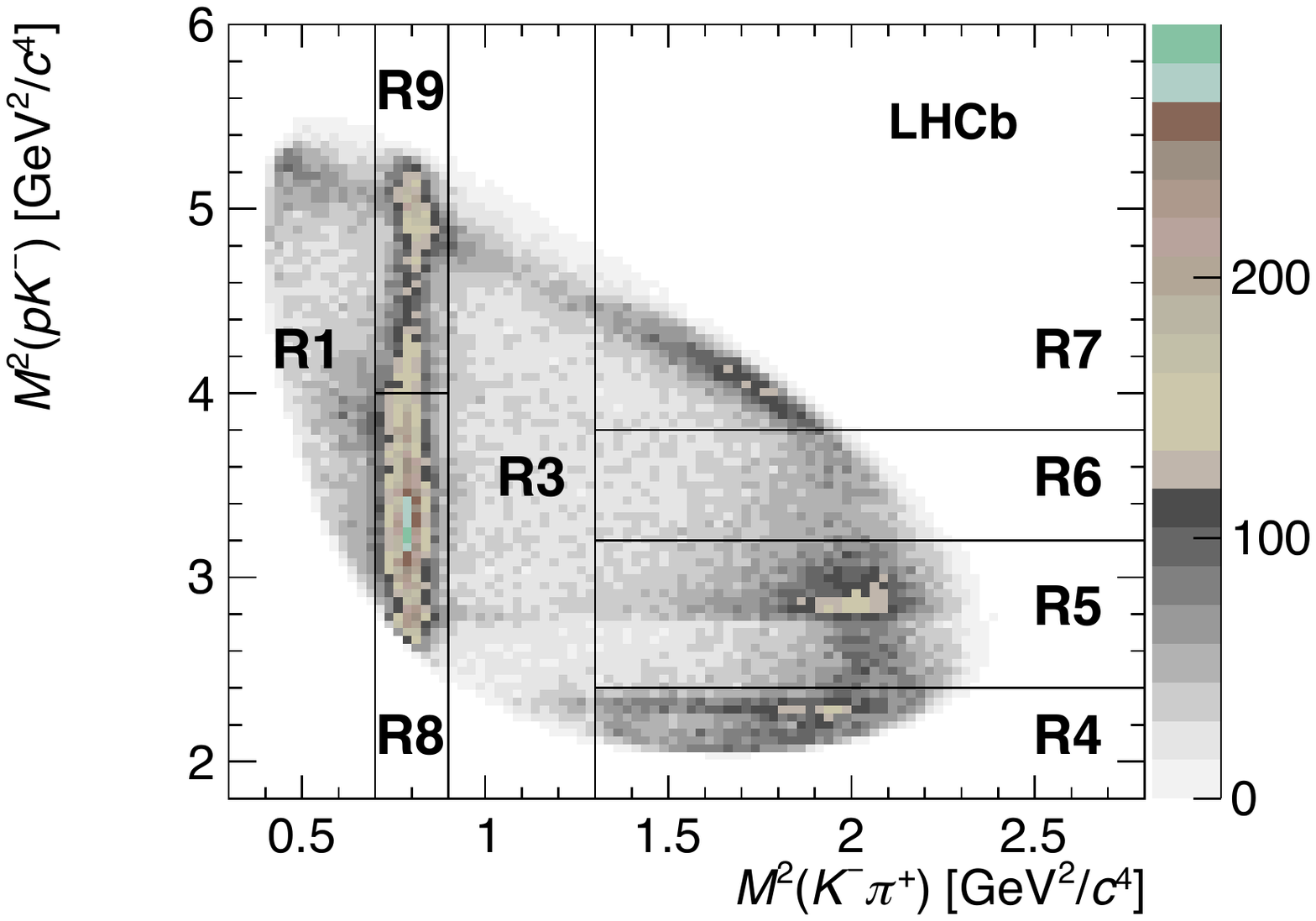}
\end{center}
\vspace*{-0.2cm}
\caption{Definition of the Dalitz plot regions for (left) \LcTopKpi  and (right)
              \XicTopKpi  decays.
              Additional regions are defined by combining regions. For \LcTopKpi
              R2=R5$\cup$R6 and for \XicTopKpi
              R2=R8$\cup$R9, R10=R4$\cup$R5 and R11=R4$\cup$R5$\cup$R6$\cup$R7. The presented distributions correspond
              to the 2012 data sample.}
\label{fig:Dalitzreg}
\end{figure}

To increase the power of the kNN method,
the Dalitz plot is divided into regions
defined around the expected resonances.
It can provide one of the necessary conditions for observation of \CPV: large relative strong phases in the final states of interfering amplitudes
of the intermediate resonance states.
The Dalitz plot is partitioned 
into six regions for the decays of the  
\Lcp control mode and eleven regions for signal \Xicp decays according to the present of resonances of the phase space, as shown in Fig.~\ref{fig:Dalitzreg}. 
The definitions of the regions
are also given in Tables~\ref{tab:DalitzCFreg} and~\ref{tab:Dalitzreg}
for \Lcp and \Xicp baryons, respectively.   
For \Lcp decays the $K^*(892)$, $K^* (1430)$, $\Deltares (1232)$,
$\Lz (1520)$, $\Lz (1670)$, $\Lz (1690)$
resonances are seen in data, whilst for \Xicp decays additional
resonances are seen, namely $\Lz (1520)$, $\Lz (1600)$,
$\Lz (1710)$, $\Lz (1800)$, $\Lz (1810)$,
$\Lz (1820)$, $\Lz (1830)$, $\Lz (1890)$,
$\Deltares (1600)$, $\Deltares (1620)$ and $\Deltares (1700)$.
For \Lcp decays there are four independent
regions (R1--R4), whilst the region R2 is further split
into the high $M^2(pK^-)$ region (R6) and the low $M^2(pK^-)$
region (R5). For \Xicp
there are seven independent regions (R1--R7),
whilst the region R2 is split in mass $M^2(pK^-)$ in two regions at larger mass (R9) and
smaller mass (R8), R2=R8$\cup$R9, 
similarly for R10 and R11, where 
R10=R4$\cup$R5, and R11=R4$\cup$R5$\cup$R6$\cup$R7. 
Region R0 is the full Dalitz plot.

\begin{table}[ht]
 \caption{Definitions of the Dalitz plot regions for the control
              mode, \LcTopKpi.}
\begin{center}\begin{tabular}{l l}\hline
   Region & Definition \\ 
  \hline
   R0    & Full Dalitz plot \\
   \hline
   R1    & $M^2(K^-\pi^+)<0.7{\rm ~GeV^2}/c^4$ \\
   R2    & $0.7\leq M^2(K^-\pi^+)< 0.9{\rm ~GeV^2}/c^4$  \\
   R3    & $M^2(K^-\pi^+)\geq 0.9{\rm ~GeV^2}/c^4$, $M^2(pK^-)< 2.8{\rm ~GeV^2}/c^4$ \\
   R4    & $M^2(K^-\pi^+)\geq 0.9{\rm ~GeV^2}/c^4$, $M^2(pK^-)\geq 2.8{\rm ~GeV^2}/c^4$ \\
   \hline
   R5    & $0.7\leq M^2(K^-\pi^+)< 0.9{\rm ~GeV^2}/c^4$, $M^2(pK^-)<3.2{\rm ~GeV^2}/c^4$ \\
   R6    & $0.7\leq M^2(K^-\pi^+)< 0.9{\rm ~GeV^2}/c^4$, $M^2(pK^-)\geq 3.2{\rm ~GeV^2}/c^4$ \\
   \hline
  \end{tabular}\end{center}
\label{tab:DalitzCFreg}
\end{table}

\begin{table}[ht]
 \caption{Definitions of the Dalitz plot regions for
               \XicTopKpi decays.}
\begin{center}\begin{tabular}{l l}\hline
   Region & Definition \\ 
  \hline
   R0    & Full Dalitz plot \\
   \hline
   R1    & $M^2(K^-\pi^+)<0.7{\rm ~GeV^2}/c^4$ \\
   R2    & $0.7\leq M^2(K^-\pi^+)< 0.9{\rm ~GeV^2/c^4}$  \\
   R3    & $0.9\leq M^2(K^-\pi^+)< 1.3{\rm ~GeV^2}/c^4$  \\
   R4    & $M^2(K^-\pi^+)\geq 1.3{\rm ~GeV^2}/c^4$, $M^2(pK^-)< 2.4{\rm ~GeV^2}/c^4$ \\
   R5    & $M^2(K^-\pi^+)\geq 1.3{\rm ~GeV^2}/c^4$, $2.4\leq M^2(pK^-)< 3.2{\rm ~GeV^2}/c^4$ \\
   R6    & $M^2(K^-\pi^+)\geq 1.3{\rm ~GeV^2}/c^4$, $3.2\leq M^2(pK^-)< 3.8{\rm ~GeV^2}/c^4$ \\
   R7    & $M^2(K^-\pi^+)\geq 1.3{\rm ~GeV^2}/c^4$, $M^2(pK^-)\geq3.8{\rm ~GeV^2}/c^4$  \\
   \hline
   R8    & $0.7\leq M^2(K^-\pi^+)< 0.9{\rm ~GeV^2}/c^4$, $M^2(pK^-)< 4{\rm ~GeV^2}/c^4$  \\
   R9    & $0.7\leq M^2(K^-\pi^+)< 0.9{\rm ~GeV^2}/c^4$, $M^2(pK^-)\geq 4{\rm ~GeV^2}/c^4$  \\
   \hline
   R10  & $M^2(K^-\pi^+)\geq 1.3{\rm ~GeV^2}/c^4$, $M^2(pK^-)< 3.2{\rm ~GeV^2}/c^4$  \\
   \hline
   R11  & $M^2(K^-\pi^+)\geq 1.3{\rm ~GeV^2}/c^4$ \\
   \hline
  \end{tabular}\end{center}
\label{tab:Dalitzreg}
\end{table}

%% file: control.tex
\section{Control mode, background and sensitivity studies}

\label{sec:control}

The $S_{\CP}$ and kNN methods are tested
using the \LcTopKpi control mode
where the \CP asymmetry is expected
to be null~\cite{Shipsey:2006zz,Artuso:2008CPV,Bianco:2020hzf,Grossman:2019xcj,Li:2019hho,Cheng:2019ggx,Calibbi:2019bay,Chala:2019fdb,Dery:2019ysp}. 
The sidebands of \XicTopKpi
candidates in the mass regions 
$2320<M(\proton\Km\pip)<2445~{\rm MeV}/c^2$ and
$2490<M(\proton\Km\pip)<2650~{\rm MeV}/c^2$
are used to check that the background does not introduce spurious asymmetries.

\begin{figure}[tb]
\begin{center}
\includegraphics[width=0.48\linewidth]{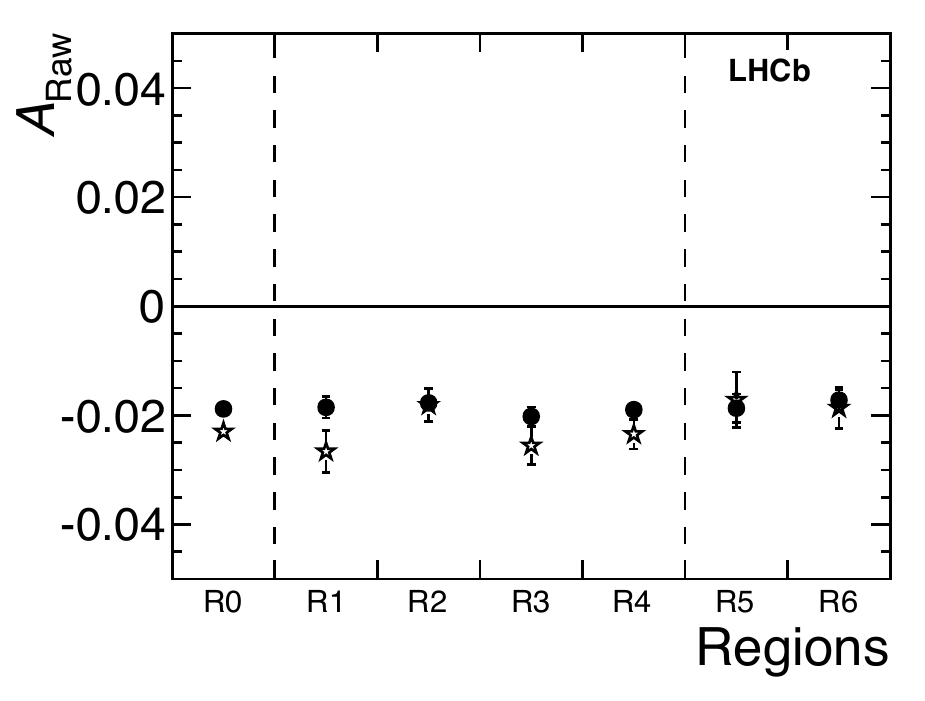}
\end{center}
\vspace*{-0.5cm}
\caption{Measured values of $A_{\rm Raw}$ in regions
               of \LcTopKpi candidate decays
               for 2011 (stars) and 2012 (dots) data samples. 
               R0 corresponds to full Dalitz plot and
               R2 is separated into R5 and R6, and these 
               regions are correlated and separated by dashed lines.}
\label{fig:arawCF}
\end{figure}

The measured total raw asymmetry is defined as 
\begin{equation}
A_{\rm Raw}=\frac{n_--n_+}{n_-+n_+},
\end{equation}
and it depends on the production asymmetry of $H_c^+$ baryons and on the detection asymmetries that arise through charge-dependent selection efficiencies due to track reconstruction, trigger selection and particle identification.
The measured value of $A_{\rm Raw}$ in each region of the Dalitz plot of \LcTopKpi decays is presented in 
Fig.~\ref{fig:arawCF}.
The measured $A_{\rm Raw}$ value integrated over the Dalitz plot
equals $-0.0230\pm 0.0016$ and $-0.0188\pm 0.0008$ in the 2011 and 2012 data samples, where the uncertainties are statistical only. Within uncertainties,
$A_{\rm Raw}$ in all regions amounts to about $-2$\%.
There is no significant difference in the 
measurement of $A_{\rm Raw}$ between the 2011 and 2012 data samples.
Since the production and detection asymmetries of \Lcp baryons
can depend on the baryon pseudorapidity, $\eta$, and \pt, the dependence of $A_{\rm Raw}$
in regions of the Dalitz plot is checked in bins of $\eta$ and \pt of the $\Lcp$ baryon.
It is observed that the value of $A_{\rm Raw}$ globally changes from bin to bin of $\eta$ and \pt
of the \Lc candidates, but for a given bin 
of $\eta$ and \pt a constant
behaviour of $A_{\rm Raw}$ 
in regions of the Dalitz plot
is maintained.

\begin{figure}[!tb]
\begin{center}
\includegraphics[width=0.98\linewidth]{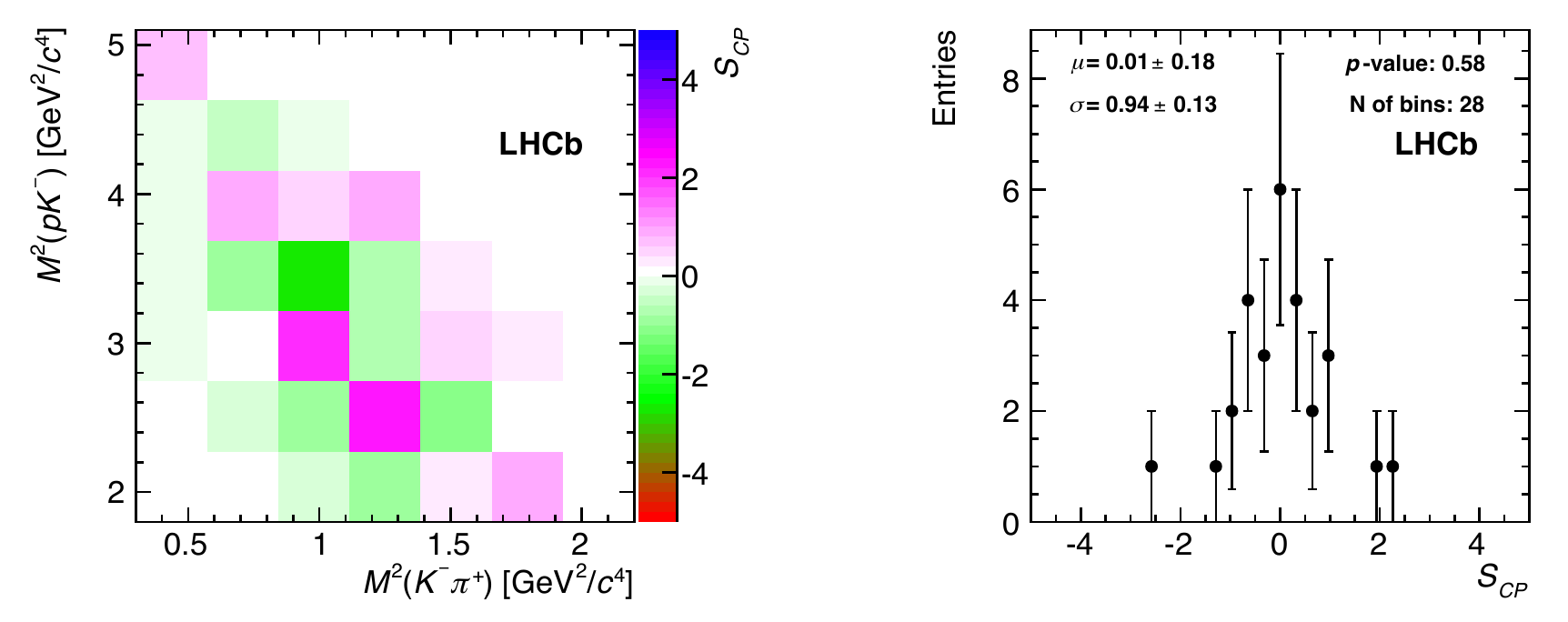}
\includegraphics[width=0.98\linewidth]{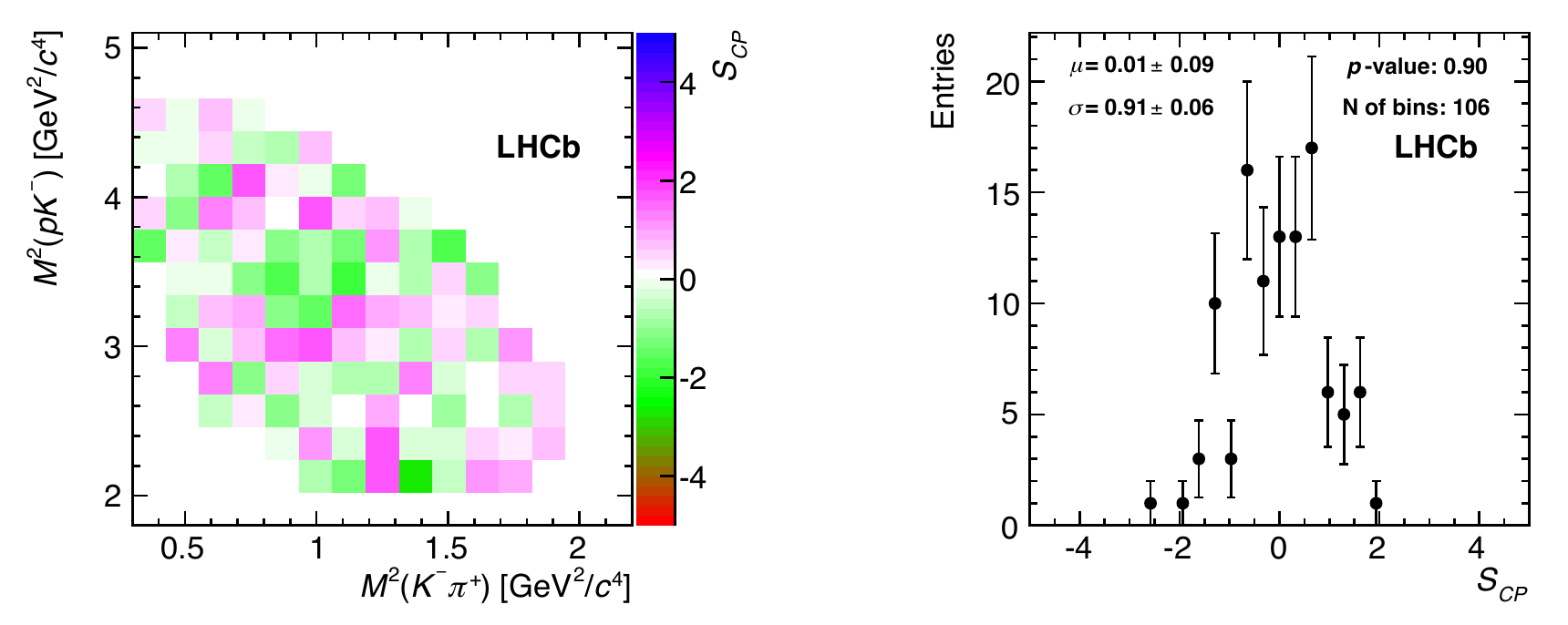}
\end{center}
\vspace*{-0.5cm}
\caption{Distributions of $S_\CP^i$ and corresponding
             one-dimensional distributions 
             for \LcTopKpi decays for the
               data collected in the 2012 data sample: 
               (top row) 28 same-size bins and 
               (bottom row) 106 same-size bins of the Dalitz plot.
               The number of analysed bins, nbins, and the 
               {\it p}-values are given.}
\label{fig:scpCF}
\end{figure}

In the $S_\CP$ method the production asymmetry and all global effects are considered by
introducing the $\alpha$ factor, following the strategy described in Sec.~\ref{sec:methodsscp}.
The {\it p}-values obtained are larger
than 58\%, consistent with the absence of localised asymmetries. As an example, Fig.~\ref{fig:scpCF} shows
the distribution of $S^i_\CP$ for \LcTopKpi decays considering
uniform binning, and for two granularities 
of the Daliz plot: 28 and 106 bins in the 2012 sample. 
Alternatively the Dalitz plot is divided
into different size bins with 
the same number of events in 
each bin. The $p$-values 
obtained are
larger than 34\%, consistent with the hypothesis 
of absence of localised asymmetries.

\begin{figure}[!tb]
\begin{center}
\includegraphics[width=0.99\linewidth]{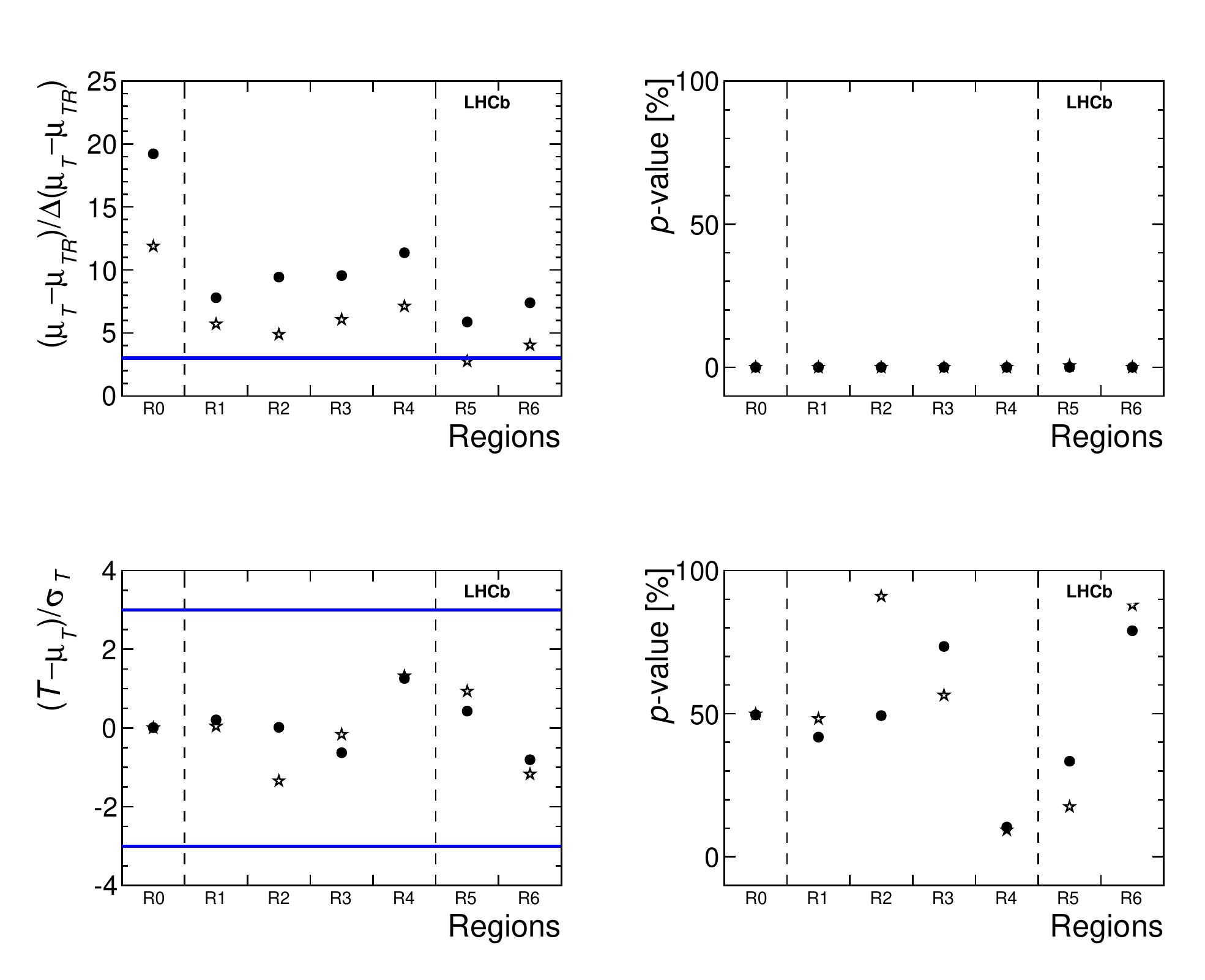}
\end{center}
\vspace*{-0.5cm}
\caption{(Top left) pulls, $(\mu_T - \mu_{TR} )/\Delta(\mu_T - \mu_{TR} )$, and (top right) the corresponding
              {\it p}-values, (bottom left) pull values of the test statistic $T$ and (bottom right) the
              corresponding {\it p}-values in regions for
              control \LcTopKpi candidate decays obtained
              using the kNN method with $n_k=50$ for data collected in
              2011 (stars) and 2012 (dots).
              The horizontal lines in the left figures represent -3 and +3 pull values.  
              R0 corresponds to full Dalitz plot and
              R2 is separated into R5 and R6, and these 
              regions are correlated and separated by dashed lines.}
\label{fig:knnCF}
\end{figure}

Following the strategy described in Sec.~\ref{sec:methodsknn},
the results of the kNN method in regions of the Dalitz plot 
for the \LcTopKpi control mode 
are presented in Fig.~\ref{fig:knnCF}, for $n_k = 50$.
The pulls,
$(\mu_T - \mu_{TR} )/\Delta(\mu_T - \mu_{TR} )$, where $\Delta(\mu_T - \mu_{TR} )$
is the statistical uncertainty on the difference
$(\mu_T - \mu_{TR} )$,
are different from zero in all regions.
The largest pull value is observed when integrated over the full Dalitz plot. 
This asymmetry is the result of the nonzero production asymmetry that is presented in Fig.~\ref{fig:arawCF} and discussed above. Pulls of the test statistic $T$,  
$((T - \mu_T )/\sigma_T )$, vary within $-3$ and $+3$, consistent
with the hypothesis of absence of localised asymmetries in any region.
The difference among data-taking years are consistent 
with statistical fluctuations.
Figure~\ref{fig:knnCF} illustrates how the larger 2012 data sample
improves the power of the kNN method.
In Run 2 (years of data taking 2016, 2017 and 2018) 
the yield is expected to be about three times larger
than that from Run 1.

The interaction cross-section of charged hadrons
with matter depends on the charged hadron momentum. 
As such, the detection asymmetries of the proton and 
kaon-pion systems are momentum dependent.
Pseudoexperiments are performed to
check whether the detection asymmetries related to particles reconstructed in the final state can generate 
a spurious \CP asymmetry. The proton detection asymmetry
varies from about 5\%
at low momentum to 1\% at $100~{\rm GeV}/c$ and is estimated using
simulations.
The kaon-pion detection asymmetry is measured
to vary from $-1.4$\% at low momentum
to $-0.7$\% at $60~{\rm GeV}/c$~\cite{LHCb-PAPER-2014-013}.
The combined effect of the two asymmetries
is found to cancel approximately and does not generate a spurious \CP
asymmetry in the Dalitz plot.

These studies are repeated 
using the candidates in the sideband of the \XicTopKpi mass 
distribution. No spurious \CP asymmetry is found for both  methods.
For further cross-checks, the control samples are divided according to the
polarity of the magnetic field. The {\it p}-values are 
found to be distributed uniformly.

The expected statistical powers of both methods are obtained by performing pseudoexperiments. 
One hundred samples of \XicTopKpi decays are generated, each with a yield and purity equivalent to that observed in 
the combined 2011 and 2012 data samples, resulting in 200\,000 \Xicp decays generated in each pseudoexperiment. 
In this model, the two-dimensional Dalitz plots are generated assuming that the \Xicp baryons are produced unpolarised. 
This model is built by including the resonances
observed in the data, using the same software 
as in Ref.~\cite{LHCb-PAPER-2016-061}. 
The same resonances as described in Sec.~\ref{sec:methodsknn} are included.
The statistical powers of the two methods are found to be comparable. 
Both methods are sensitive to a 5\% \CP asymmetry in the $K^*(892)$
and $\Deltares (1232)$ resonance regions with 3 and 5 sigma 
significances that would be observed in 69\% and 10\% of the cases for 
the kNN method and 17\% and 10\% of the cases for the $S_\CP$ method, respectively.

%% file: results.tex
\section{Results}

\subsection{Binned \boldmath $S_\CP$ method}

\begin{figure}[!tb]
\begin{center}
\includegraphics[width=0.98\linewidth]{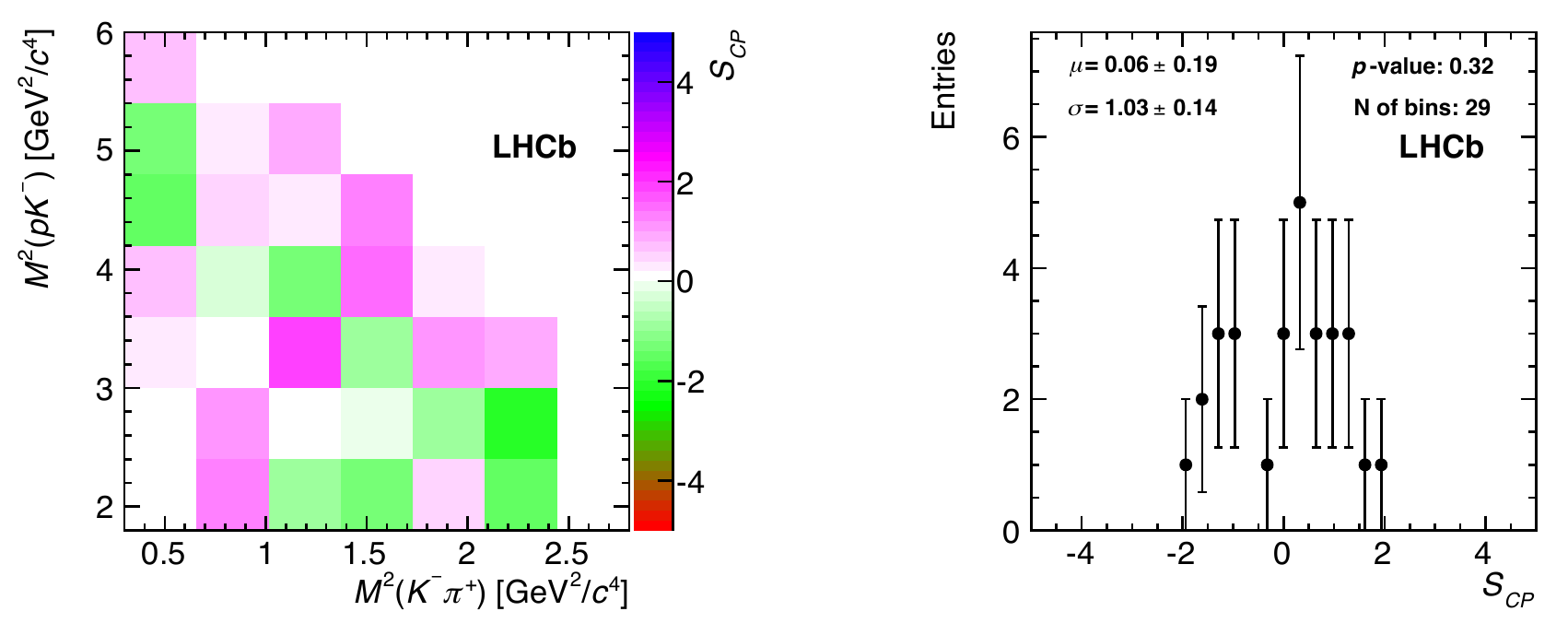}
\includegraphics[width=0.98\linewidth]{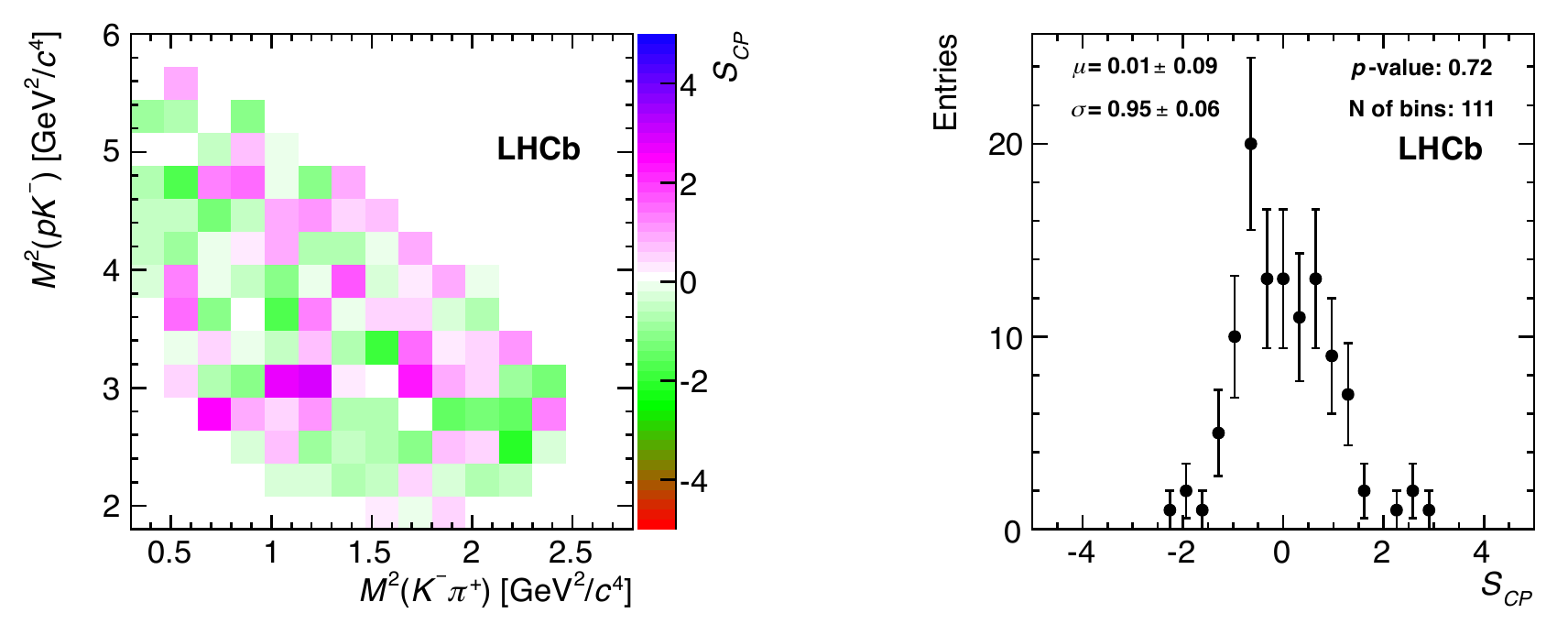}
\end{center}
\vspace*{-0.5cm}
\caption{Distributions of $S_\CP^i$ and corresponding
             one-dimensional distributions 
             for \XicTopKpi decays for the
               combined data collected 2011 and 2012:
               (top row) 29 uniform bins and 
               (bottom row) 111 uniform bins of the Dalitz plot.
               The number of analysed bins and the 
               {\it p}-values are given.}
\label{fig:scp}
\end{figure}

The binned $S_\CP$ method is applied
to look for local \CP asymmetries in  \XicTopKpi decays
following the strategy described in Sec.~\ref{sec:methodsscp}. 
The distribution of $S_\CP^i$ for \XicTopKpi 
decays considering
uniform binning, and for two granularities of the Daliz plot: 29 and 111 bins are shown
in Fig.~\ref{fig:scp} for the 
combined 2011 and 2012
data samples. 
The normalization factor $\alpha$, defined in Eq.~\ref{eq:scp},
is determined to be $1.029\pm 0.004$. 
The measured {\it p}-values using a $\chi^2$ test are larger than 32\%, 
consistent with no evidence for \CPV.  
The obtained
$S_\CP$ distributions agree with 
a normal distribution.
It is also checked that the results in the 
2011 and 2012 data samples are consistent with each other.

\subsection{Unbinned kNN method}

\begin{figure}[!tb]
\begin{center}
\includegraphics[width=0.99\linewidth]{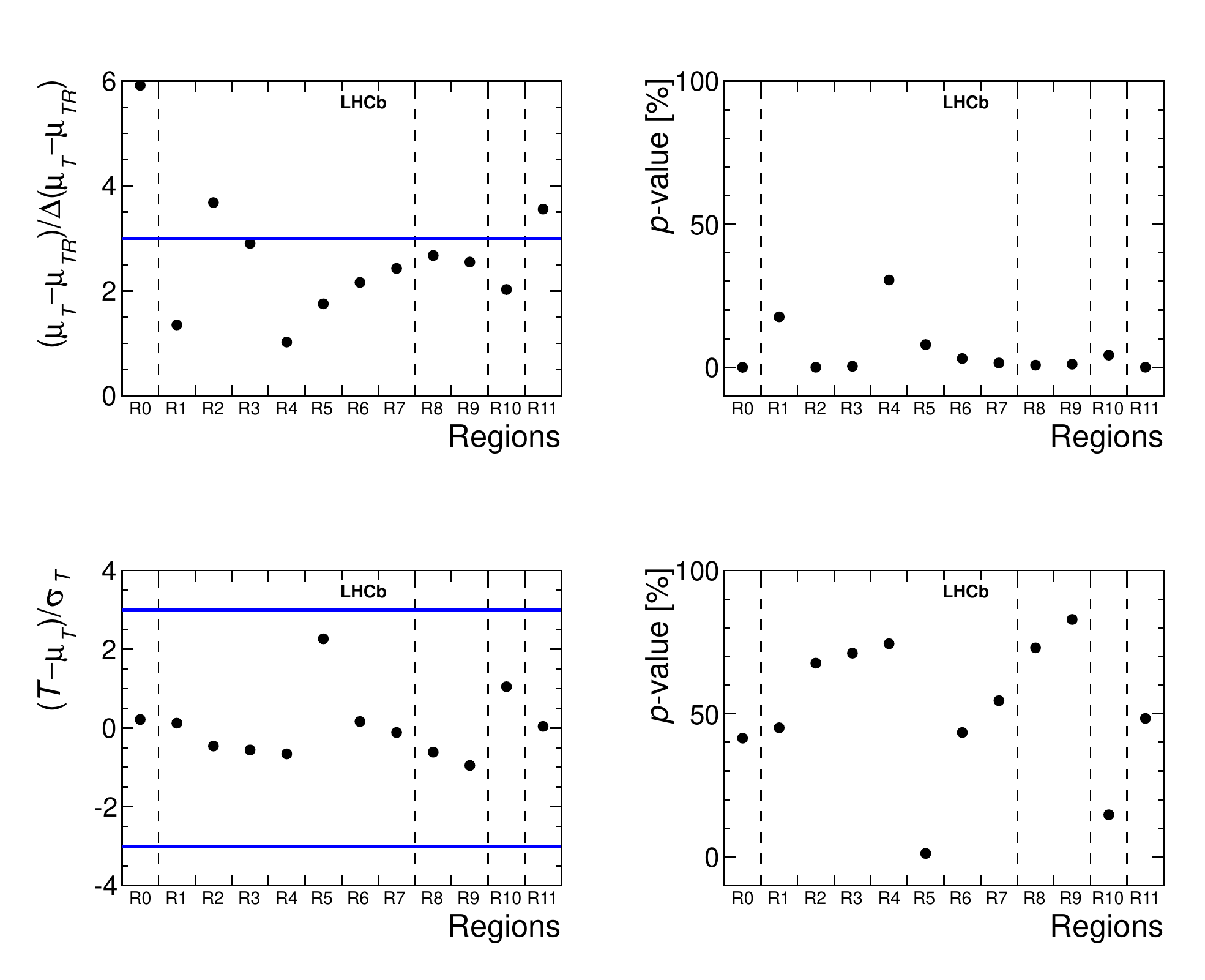}
\end{center}
\vspace*{-0.5cm}
\caption{(Top left) pulls, $(\mu_T - \mu_{TR} )/\Delta(\mu_T - \mu_{TR} )$, and (top right) the corresponding
              {\it p}-values;
              (bottom left) pull values of the test statistic $T$ and (bottom right) the
              corresponding {\it p}-values in regions for signal \XicTopKpi candidate decays obtained
              using the kNN method with $n_k=50$ for 
              combined data collected 2011 and 2012.
              The horizontal lines in the left figures represent $-3$ and $+3$ pull values.  
              R0 corresponds to full Dalitz plot and
              R2 is separated into R8 and R9, 
              R10 is separated into R4 and R5,
              R11 is separated into R4, R5, R6 and R7, and these 
              regions are correlated and separated by dashed lines.}
\label{fig:knn}
\end{figure}

\begin{figure}[htb]
\begin{center}
\includegraphics[width=0.48\linewidth]{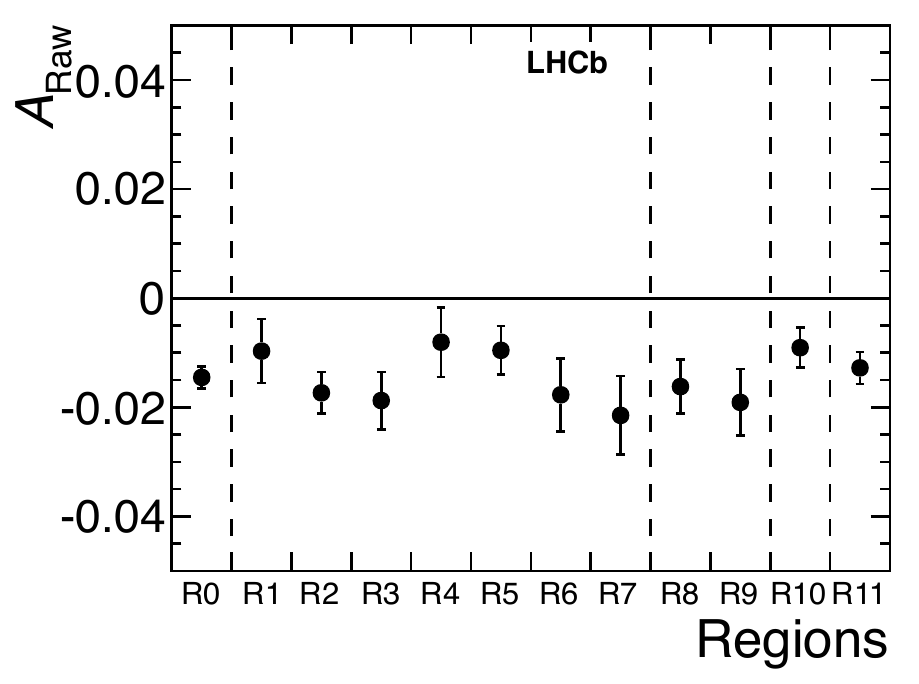}
\end{center}
\vspace*{-0.5cm}
\caption{The measured $A_{\rm Raw}$ in regions
               in signal \XicTopKpi candidate decays
               for the combined data collected in 2011 and 2012. 
               R0 corresponds to full Dalitz plot and
               R2 is separated into R8 and R9, 
               R10 is separated into R4 and R5,
               R11 is separated into R4, R5, R6 and R7, and these 
               regions are correlated and separated by dashed lines.}
\label{fig:araw}
\end{figure}

The unbinned kNN method is applied to look for \CP
asymmetry in \XicTopKpi decays, following the strategy described 
in Sec.~\ref{sec:methodsknn}. 
The results are presented 
in Fig.~\ref{fig:knn} for $n_k = 50$
for the merged 2011 and 2012 data samples.
The measured pull values, $((\mu_T - \mu_{TR} )/\Delta(\mu_T - \mu_{TR} ))$, are different from zero.
The largest value of pull is observed integrated over the full Dalitz plot. 
This is due to  the expected
nonzero production and detector asymmetries, that is presented in Fig.~\ref{fig:araw}.
The measured $A_{\rm Raw}$ is 
constant within uncertainties in all regions.

The pulls of the test statistic $T$, 
$((T - \mu_T )/\sigma_T )$, shown in Fig.~\ref{fig:knn} vary 
within $-3$ and $+3$,
consistent with the hypothesis 
of absence of localised asymmetries. 
To check for any systematic effects the kNN test is
repeated for the individual 2011 and 2012 data samples as well as for samples separated 
according to the polarity of the magnetic field. 
All obtained results are compatible within uncertainties and no systematic effects are observed.

\begin{figure}[!htb]
\vspace*{-2cm}
\begin{center}
\includegraphics[width=0.99\linewidth]{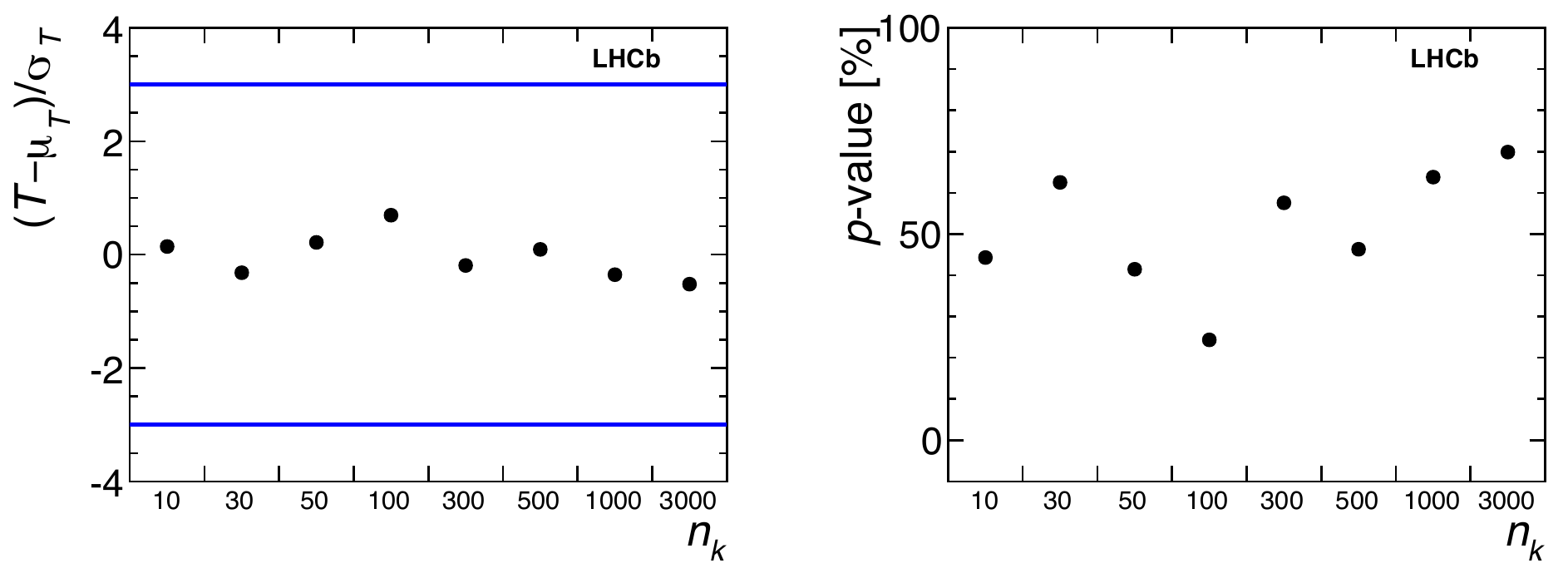}
\end{center}
\vspace*{-0.5cm}
\caption{ (Left) the pull values of the test statistic $T$ and (right) the corresponding {\it p}-value 
              dependence on the $n_k$ parameter for the whole Dalitz plot (region R0)
              for \XicTopKpi candidate decays obtained
              using the kNN method for the combined data collected in 2011 and 2012.
              The horizontal lines in the left figures represent $-3$ and $+3$ pull values.  
              The points are determined with different $n_k$
              using same data sample, therefore are correlated.}
\label{fig:nkdependence}
\end{figure}

Since the sensitivity of the method can depend 
on the $n_k$ parameter, the analysis is repeated with different values of $n_k$ from
10 up to 3000. Only $T$ and $\sigma_T$ depend on  $n_k$. Pulls of the statistic $T$ for the entire Dalitz plot
are shown in Fig.~\ref{fig:nkdependence}.
All results show no significant deviation from the hypothesis of \CP
symmetry.

%% file: summary.tex
\section{Conclusions}

Model-independent searches for \CP
violation in \XicTopKpi decays
are presented using the binned $S_\CP$ 
and the unbinned kNN
methods. The \LcTopKpi 
candidates and the sideband regions of
\XicTopKpi candidates are used to ensure that no spurious charge asymmetries affect the methods. 
Both methods are sensitive to  \CP asymmetry larger than a 5\% in the regions
around the $K^*(892)$ and the $\Deltares$(1232).
The obtained results are consistent with the  absence of \CP violation in \XicTopKpi decays.

%% file: acknowledgements.tex
\section*{Acknowledgements}
%
%
\noindent We express our gratitude to our colleagues in the CERN
accelerator departments for the excellent performance of the LHC. We
thank the technical and administrative staff at the LHCb
institutes.
We acknowledge support from CERN and from the national agencies:
CAPES, CNPq, FAPERJ and FINEP (Brazil); 
MOST and NSFC (China); 
CNRS/IN2P3 (France); 
BMBF, DFG and MPG (Germany); 
INFN (Italy); 
NWO (Netherlands); 
MNiSW and NCN (Poland); 
MEN/IFA (Romania); 
MSHE (Russia); 
MinECo (Spain); 
SNSF and SER (Switzerland); 
NASU (Ukraine); 
STFC (United Kingdom); 
NSF (USA).
We acknowledge the computing resources that are provided by CERN, IN2P3
(France), KIT and DESY (Germany), INFN (Italy), SURF (Netherlands),
PIC (Spain), GridPP (United Kingdom), RRCKI and Yandex
LLC (Russia), CSCS (Switzerland), IFIN-HH (Romania), CBPF (Brazil),
PL-GRID (Poland) and OSC (USA).
We are indebted to the communities behind the multiple open-source
software packages on which we depend.
Individual groups or members have received support from
AvH Foundation (Germany);
EPLANET, Marie Sk\l{}odowska-Curie Actions and ERC (European Union);
ANR, Labex P2IO and OCEVU, and R\'{e}gion Auvergne-Rh\^{o}ne-Alpes (France);
Key Research Program of Frontier Sciences of CAS, CAS PIFI, and the Thousand Talents Program (China);
RFBR, RSF and Yandex LLC (Russia);
GVA, XuntaGal and GENCAT (Spain);
the Royal Society
and the Leverhulme Trust (United Kingdom);
Laboratory Directed Research and Development program of LANL (USA).

%% file: LHCb_Authorship_25-Jun-2019.tex
\centerline
{\large\bf LHCb collaboration}
\begin
{flushleft}
\small
R.~Aaij$^{31}$,
C.~Abell{\'a}n~Beteta$^{49}$,
T.~Ackernley$^{59}$,
B.~Adeva$^{45}$,
M.~Adinolfi$^{53}$,
H.~Afsharnia$^{9}$,
C.A.~Aidala$^{79}$,
S.~Aiola$^{25}$,
Z.~Ajaltouni$^{9}$,
S.~Akar$^{64}$
% ...

{\footnotesize \it

% ...
$ ^{9}$Universit{\'e} Clermont Auvergne, CNRS/IN2P3, LPC, Clermont-Ferrand, France\\
% ...
$ ^{25}$INFN Sezione di Milano, Milano, Italy\\
% ...
$ ^{31}$Nikhef National Institute for Subatomic Physics, Amsterdam, Netherlands\\
% ...
$ ^{45}$Instituto Galego de F{\'\i}sica de Altas Enerx{\'\i}as (IGFAE), Universidade de Santiago de Compostela, Santiago de Compostela, Spain\\
% ...
$ ^{49}$Physik-Institut, Universit{\"a}t Z{\"u}rich, Z{\"u}rich, Switzerland\\
% ...
$ ^{53}$H.H. Wills Physics Laboratory, University of Bristol, Bristol, United Kingdom\\
% ...
$ ^{59}$Oliver Lodge Laboratory, University of Liverpool, Liverpool, United Kingdom\\
% ...
$ ^{64}$University of Cincinnati, Cincinnati, OH, United States\\
% ...
$ ^{67}$Syracuse University, Syracuse, NY, United States\\
% ...
$ ^{79}$University of Michigan, Ann Arbor, United States, associated to $^{67}$\\
\bigskip
% ...
}
\end{flushleft}